\documentstyle[aps,eqsecnum,floats]{revtex}
\input{psfig.tex}
\begin{document}
\draft
\preprint{\vbox{\hbox{CU-TP-787} 
                \hbox{CAL-617}
                \hbox{CfA-????}
		\hbox{FERMILAB-Pub-96/???-A}
                \hbox{astro-ph/9611125}
}}
\title{Statistics of Cosmic Microwave Background Polarization}
\author{Marc Kamionkowski\footnote{kamion@phys.columbia.edu}}
\address{Department of Physics, Columbia University, 538 West
120th Street,
New York, New York~~10027}
\author{Arthur Kosowsky\footnote{akosowsky@cfa.harvard.edu}}
\address{Harvard-Smithsonian Center for Astrophysics,
60 Garden Street, Cambridge, Massachusetts~~02138
\\and\\
Department of Physics, Lyman Laboratory, Harvard University,
Cambridge, Massachusetts~~02138}
\author{Albert Stebbins\footnote{stebbins@fnal.gov}}
\address{NASA/Fermilab Astrophysics Center, Fermi National
Accelerator Laboratory, Batavia, IL 60510-0500}
\maketitle

\begin{abstract}
We present a formalism for analyzing a full-sky temperature
and polarization map of the cosmic microwave background.
Temperature maps are analyzed by expanding over
the set of spherical harmonics to give multipole moments
of the two-point correlation function. Polarization, which is
described by a second-rank tensor, can be treated analogously
by expanding in the appropriate tensor spherical harmonics. We
provide expressions for the complete
set of temperature and polarization multipole moments for 
scalar and tensor metric perturbations. Four sets of
multipole moments completely describe isotropic temperature
and polarization correlations; for scalar metric 
perturbations one set is identically zero, giving the 
possibility of a clean determination of the vector and tensor 
contributions. The variance with which the multipole moments
can be measured in idealized experiments is evaluated, including
the effects of detector noise, sky coverage, and beam width. Finally,
we construct coordinate-independent polarization two-point 
correlation functions, express them
in terms of the multipole moments, and derive small-angle limits.

\end{abstract}


\def\hatn{{\bf \hat n}}
\def\hatnprime{{\bf \hat n'}}
\def\hatnone{{\bf \hat n}_1}
\def\hatntwo{{\bf \hat n}_2}
\def\hatni{{\bf \hat n}_i}
\def\hatnj{{\bf \hat n}_j}
\def\vecx{{\bf x}}
\def\veck{{\bf k}}
\def\hatx{{\bf \hat x}}
\def\hatk{{\bf \hat k}}
\def\hatz{{\bf \hat z}}
\def\VEV#1{{\left\langle #1 \right\rangle}}
\def\cP{{\cal P}}
\def\noise{{\rm noise}}
\def\pix{{\rm pix}}
\def\map{{\rm map}}

\section{Introduction}

With the advent of a new generation of balloon-borne and
ground-based experiments \cite{whitereview} and satellite
missions \cite{mapsatellite,cobrassamba}, the cosmic
microwave background (CMB) will provide an unprecedented window
to the early Universe.  In addition to determining the origin of
large-scale structure, it has been argued that CMB temperature
maps may determine cosmological parameters and the ionization
history of the Universe, and perhaps probe long-wavelength
gravitational waves 
\cite{bondetal,acoustic,kamspergelsug,parameters,newparameters,neutrinos}.

Any mechanism which produces temperature anisotropies
will invariably lead to polarization as well
\cite{be87,zaldharari,tensors,annals,flatsky}.  
Temperature fluctuations are the result of perturbations in
the gravitational potentials, which contribute directly to
the fluctuations via gravitational redshifting (the Sachs-Wolfe
effect \cite{sachswolfe}) and which drive acoustic oscillations of the
primordial plasma \cite{acoustic}. These processes result in
temperature fluctuations which are the same order of
magnitude as the metric perturbations. In contrast, polarization
is not directly generated by metric perturbations: a net
polarization arises from Compton scattering only when the
incident radiation field possesses a non-zero quadrupole moment 
\cite{cct95,annals},
but only monopole and dipole fluctuations are possible as long
as the photons in the Universe remain tightly coupled to
the charged electrons. Polarization is only generated very near
the surface of last scattering as the photons begin to decouple
{}from the electrons and generate a quadrupole moment through
free-streaming \cite{kl96}. Since by this time most of the electrons have
recombined into neutral hydrogen, the number of scatterers
available to produce polarization is reduced, so
CMB polarization fluctuations are characteristically at 
a part in $10^6$, an order of magnitude below
the temperature fluctuations.

A polarization map will provide information that complements that from a
temperature map.  For example, polarization may help distinguish the
gravitational-potential and peculiar-velocity contributions
to the acoustic peaks in the
temperature-anisotropy power spectrum \cite{zaldharari}.  
In models with reionization, some of the information lost from
damping of temperature anisotropies will be regained in the
polarization spectrum \cite{reionization}.  Perhaps most importantly, the
density-perturbation and gravitational-wave or vorticity
contributions to the anisotropy can be geometrically decomposed
with a polarization map \cite{cct95,letter,sz96b,sz96c}.
Furthermore, although these non-scalar signals are expected to be
small, they will not be swamped by cosmic variance from scalar
modes (as discussed further below).  Detection of gravity waves
is important for testing inflation and for learning about the
inflaton potential which drove inflation \cite{inflationtest}.

Realistically, detection will present a
significant experimental challenge.  Current results limit the
magnitude of linear polarization to roughly a part in $10^5$
\cite{currentlimits}.  Experiments being planned
or built will improve sensitivities by at least an order
of magnitude \cite{newexperiments}. 
The MAP satellite will make polarized measurements of the
entire microwave sky in around a million pixels with a
precision of around one part in $10^5$ per pixel \cite{mapsatellite}. 
If CMB polarization is not discovered by a ground or
balloon experiment in
the next four years, this satellite will almost certainly
make the first detection. The COBRAS/SAMBA satellite may also
make polarized measurements \cite{cobrassamba}.  These
experimental prospects, as well as the theoretical considerations
above, motivate the analysis presented in this paper. 

Previous theoretical treatments of CMB polarization have
relied on a small-angle approximation, which is valid
when considering patches of the sky small enough to be
approximated as flat.  Upcoming polarization
maps will require a more sophisticated formalism. In this
paper, we develop in detail a description of polarization
on the full sky. The Stokes parameters conventionally
used to describe polarization are not invariant under
rotations of the coordinate system used to describe them,
unlike temperature fluctuations,
but rather transform as a second-rank tensor \cite{annals}.
By expressing the polarization in terms of a complete,
orthonormal set of tensor basis functions on the celestial
sphere, power spectra and correlation functions which
are independent of the coordinate system can be constructed.
Earlier work on small patches of the sky chose a particular
reference coordinate system which completely defines the
polarization but obscures the physical interpretation
of the polarization pattern. 
Also, the signal from vector and tensor perturbations
is expected to contribute to CMB polarization primarily
at large angles on the sky through gravitational effects, so the correct
full-sky analysis is essential.

Our formalism is stated in terms of differential geometry on the
sphere, using a notation widely used in general relativity.  Similar
calculations have recently been performed by Seljak and Zaldarriaga
\cite{sz96b,sz96c}, using spin-weighted spherical harmonics
\cite{spinweight}.   Although the formalisms employed differ
substantially and the calculations are quite lengthy,
we have verified that the end results are equivalent where they
overlap, giving us confidence both are correct.

After a brief
review of Stokes parameters, the next Section defines the tensor
spherical harmonic basis
functions and gives useful explicit expressions and formulae for decomposing a
polarization map into its harmonic components.  Section III covers the
statistics of the expansion coefficients of the temperature and polarization
harmonics, derivations of variance estimates for the various multipole moments
in idealized experiments, and a recipe for simulating a combined polarization
and temperature map given theoretical angular power spectra.  Section IV
derives exact expressions for all of the multipole moments from scalar and
tensor metric fluctuations, expressed in terms of the conventional Fourier
components of radiation brightnesses. Section V then treats two-point
correlation functions of the Stokes parameters in a coordinate-independent
manner and expresses the multipole moments and correlation
functions in terms of each other.  We also reproduce flat-sky
results by taking small-angle limits and make an explicit
connection with earlier work in particular fixed coordinate
systems.  Finally, a summary and discussion section briefly
considers detection prospects for various polarization signals.
A pair of mathematical appendices collect results from
differential geometry on the sphere and useful identities of
Legendre polynomials and spherical harmonics.

\section{Description of Polarization}

\subsection{Review of Stokes Parameters}

The cosmic microwave background is characterized
completely by its temperature and polarization
in each direction on the sky (assuming its
frequency spectrum is a perfect blackbody).
Polarized radiation is described
in terms of the Stokes parameters $Q$, $U$, and
$V$ \cite{optics}. For a monochromatic electromagnetic wave
of frequency $\omega_0$ propagating in the
$z$-direction, the components of the wave's 
electric field vector at a given point in space
can be written as
\begin{equation}
     E_x = a_x(t) \cos[\omega_0 t - \theta_x(t)],\qquad\qquad
     E_y = a_y(t) \cos[\omega_0 t - \theta_y(t)].
\label{efield}
\end{equation}
If these two components are correlated, then the
wave is said to be polarized. The Stokes parameters
are defined as the time averages
\begin{eqnarray}
     I & \equiv & \left\langle a_x^2\right\rangle 
                         +\left\langle a_y^2\right\rangle ,\\
     Q & \equiv & \left\langle a_x^2\right\rangle 
                         -\left\langle a_y^2\right\rangle ,\\
     U & \equiv & \left\langle 2a_x a_y \cos(\theta_x -
     \theta_y)\right\rangle ,\\ 
     V & \equiv & \left\langle 2a_x a_y \sin(\theta_x - \theta_y)\right\rangle.
\label{stokesdef}
\end{eqnarray}
The parameter $I$ gives the radiation intensity which is
positive definite. The other three parameters can take
either sign and describe the polarization state.
For unpolarized radiation, $Q=U=V=0$. The Stokes
parameters are additive for incoherent superpositions
of waves, which makes them natural variables
for describing polarized radiative transport.

	In most applications polarization is measured in units of intensity;
however it is conventional and convenient when studying the CMB to 
express polarization in terms of the difference in brightness temperature of a
particular polarization state from that of the mean brightness temperature of
the CMB.  The rationale for this convention
comes from the well-known result that the
spectrum of polarization induced in the CMB is exactly the same as a
temperature anisotropy, so in brightness temperature units the polarization
should be independent of frequency.

The Stokes parameters $I$ and $V$ describe physical
observables and are independent of the choice of
coordinate system. However, $Q$ and $U$ describe
orthogonal modes of linear polarization and
depend on the axes in relation to which the linear
polarization is defined. From Eqs.~(\ref{stokesdef}),
it is easy to show that when the coordinate system
is rotated by an angle $\alpha$, the same radiation
field is now described by the parameters
\begin{eqnarray}
     Q' &=& Q\cos(2\alpha) + U\sin(2\alpha), \nonumber \\
     U' &=& -Q\sin(2\alpha) + U\cos(2\alpha).
\label{stokestransf}
\end{eqnarray}
Stated another way, under rotations of the coordinate
system around the direction of propagation,
the $Q$ and $U$ Stokes parameters transform
like the independent components of a two-dimensional, second rank
symmetric trace-free (STF) tensor. Thus we can equally well describe
the linear polarization state by a polarization tensor
${\cal P}_{ab}$, which coincides with the photon
density matrix \cite{annals}.

\subsection{Scalar and Tensor Harmonic Expansions}

Suppose we have an all-sky map of the CMB temperature
$T({\bf \hat n})$ and polarization tensor
${\cal P}_{ab}({\bf \hat n})$.  The polarization tensor is a $2\times2$
symmetric (${\cal P}_{ab}={\cal P}_{ba}$) and trace-free ($g^{ab}
{\cal P}_{ab}=0$) tensor, so it is specified by two real
quantities.  Given the Stokes parameters $Q$ and $U$
measured in any coordinate system, we can construct ${\cal
P}_{ab}$.  For example, in spherical polar coordinates,
($\theta$,$\phi$), the metric is $g_{ab}={\rm
diag}(1,\sin^2\theta)$ and
\begin{equation}
  {\cal P}_{ab}(\hatn)={1\over 2} \left( \begin{array}{cc}
   Q(\hatn) & -U(\hatn) \sin\theta \\
   \noalign{\vskip6pt}
   - U(\hatn)\sin\theta & -Q(\hatn)\sin^2\theta \\
   \end{array} \right).
\label{whatPis}
\end{equation}
The factors of $\sin\theta$ must be included since the coordinate basis for 
$(\theta,\phi)$ is an orthogonal, but not an orthonormal basis.
(For more details of differential geometry on the 2-sphere, see
Appendix A.) The Compton scattering process which thermalizes
the CMB and generates polarization cannot produce any net circular
polarization \cite{chandra}; 
thus we expect $V=0$ for the microwave background
and do not consider the $V$ Stokes parameter further.
Note the spherical polar coordinate system adopted in this paper
gives an outward direction for the $z$-axis, which is opposite the
radiation propagation direction. The convention with the $z$-axis
in the direction of propagation is sometimes used, particularly
in Ref.~\cite{annals}; this leads to the opposite
sign for the $U$ Stokes parameter, but all results are unchanged.

In the usual way, we can expand the temperature pattern
$T(\hatn)$ in a set of
complete orthonormal basis functions, the spherical harmonics:
\begin{equation}
     {T(\hatn) \over T_0}=1+\sum_{l=1}^\infty\sum_{m=-l}^l
     a^{\rm T}_{(lm)}\,Y_{(lm)}(\hatn)
\label{Texpansion}
\end{equation}
where
\begin{equation}
 a^{\rm T}_{(lm)}={1\over T_0}\int d\hatn\,T(\hatn) Y_{(lm)}^*(\hatn)
\label{temperaturemoments}
\end{equation}
are the temperature multipole coefficients and $T_0$ is the
mean CMB temperature. The $l=1$ term in Eq.~(\ref{Texpansion}) is
indistinguishable from the kinematic dipole and is normally ignored.

Similarly, we can expand the polarization tensor in terms of a complete set of
orthonormal basis functions for symmetric trace-free $2\times2$ tensors on the
2-sphere, 
\begin{equation}
     {{\cal P}_{ab}(\hatn)\over T_0} = \sum_{l=2}^\infty\sum_{m=-l}^l
     \left[ a_{(lm)}^{\rm G}
     Y_{(lm)ab}^{\rm G}(\hatn) + a_{(lm)}^{\rm C} Y_{(lm)ab}^{\rm C}
     (\hatn) \right],
\label{Pexpansion}
\end{equation}
where the expansion coefficients are given by
\begin{equation}
     a_{(lm)}^{\rm G}={1\over T_0}\int \, d\hatn {\cal P}_{ab}(\hatn)
                             Y_{(lm)}^{{\rm G} \,ab\, *}(\hatn), \qquad\qquad
     a_{(lm)}^{\rm C}={1\over T_0}\int d\hatn\, {\cal P}_{ab}(\hatn)
                                      Y_{(lm)}^{{\rm C} \, ab\, *}(\hatn),
\label{defmoments}
\end{equation}
which follow from the orthonormality properties
\begin{equation}
 \int d\hatn\,Y_{(lm)ab}^{{\rm G}\,*}(\hatn)\,Y_{(l'm')}^{{\rm G}\,\,ab}(\hatn)
=\int d\hatn\,Y_{(lm)ab}^{{\rm C}\,*}(\hatn)\,Y_{(l'm')}^{{\rm C}\,\,ab}(\hatn)
=\delta_{ll'} \delta_{mm'},\nonumber
\end{equation}
\begin{equation}
\int d\hatn\,Y_{(lm)ab}^{{\rm G}\, *}(\hatn)\,
Y_{(l'm')}^{{\rm C}\,\,ab}(\hatn)
=0.
\label{norms}
\end{equation}
Note that unlike scalar harmonics, the tensor harmonics only exist for $l\ge2$
\cite{albert}.

The basis functions 
$Y_{(lm)ab}^{\rm G}(\hatn)$ and $Y_{(lm)ab}^{\rm C}(\hatn)$
are given in terms of
covariant derivatives of the spherical harmonics by \cite{albert}
\begin{equation}
     Y_{(lm)ab}^{\rm G} = N_l
     \left( Y_{(lm):ab} - {1\over2} g_{ab} Y_{(lm):c}{}^c \right),
\label{Yplusdefn}
\end{equation}
and
\begin{equation}
     Y_{(lm)ab}^{\rm C} = { N_l \over 2}    
     \left(\vphantom{1\over 2} 
       Y_{(lm):ac} \epsilon^c{}_b +Y_{(lm):bc} \epsilon^c{}_a \right),
\label{Ytimesdefn}
\end{equation}
where $\epsilon_{ab}$ is the completely antisymmetric tensor,
the ``:'' denotes covariant differentiation on the 2-sphere,
and
\begin{equation}
     N_l \equiv \sqrt{ {2 (l-2)! \over (l+2)!}}
\label{Nleqn}
\end{equation}
is a normalization factor.

The existence of two sets of basis functions,
labeled here by ``G'' and
``C'', is due to the fact that an STF
$2\times2$ tensor is specified by two independent parameters.
In two dimensions, any STF tensor can be uniquely decomposed into a
part of the form $A_{:ab}-(1/2)g_{ab} A_{:c}{}^c$ and another part of the form
$B_{:ac}\epsilon^c{}_b+B_{:bc}\epsilon^c{}_a$ where $A$ and $B$ are two scalar
functions.   This decomposition is quite similar to the decomposition of a
vector field into a part which is the gradient of a scalar field 
and a part which is the curl of a vector field; hence we use the notation
G for ``gradient'' and C for
``curl''.  Since the $Y_{(lm)}$'s provide a complete basis for scalar
functions on the sphere, the $Y_{(lm)ab}^{\rm G}$ and
$Y_{(lm)ab}^{\rm C}$ tensors provide a complete basis for G-type
and C-type STF tensors, respectively.  This G/C decomposition is
also known as the scalar/pseudo-scalar decomposition \cite{albert}.

Incidentally, these tensor spherical harmonics are identical to
those which appear in the theory of gravitational radiation 
\cite{thorne,zerilli}.  The propagating degrees of freedom of
gravitational field perturbations
are described by a spin-2 tensor.  Computing
the flux of gravitational radiation from a source requires
the components of the gravitational field tangent to a
sphere around the source which are induced by the motions of that
source.  Our G harmonics are often \cite{thorne}---but not
always \cite{zerilli}---referred to as having ``electric-type''
parity, since an electric field can be written as the gradient
of a scalar.  Likewise, our C harmonics have ``magnetic-type''
parity since they are the curl of a vector field. The two varieties
of harmonics also correspond to electric and magnetic multipole
radiation.

Integration by parts
transforms Eqs.~(\ref{defmoments})
into integrals over scalar spherical
harmonics and derivatives of the polarization tensor:
\begin{equation}
     a_{(lm)}^{\rm G} = {N_l\over T_0}
     \int d\hatn \, Y_{(lm)}^*(\hatn)\,
     {\cal P}_{ab}{}^{:ab}(\hatn),
\label{Gmomentseasy}
\end{equation}
\begin{equation}
     a_{(lm)}^{\rm C} = {N_l\over T_0} 
     \int d\hatn \, Y_{(lm)}^*(\hatn)\,
     {\cal P}_{ab}{}^{:ac}(\hatn) \epsilon_c{}^b,
\label{Cmomentseasy}
\end{equation}
where the second equation uses the fact that
$\epsilon^{ab}{}_{:c}=0$.  These forms are useful for theoretical calculations
of the multipole moments.  We don't recommend taking second derivatives of real
data!  Since $T$ and ${\cal P}_{ab}$ are real, all of
the multipole must obey the reality condition
\begin{equation}
     a_{(lm)}^{\rm X\,*} =
     (-1)^m a_{(l,-m)}^{\rm X}.
\label{almsymmetry}
\end{equation}
where ${\rm X}= \{{\rm T,G,C}\}$.

\subsection{Explicit Form of the Harmonics}

In spherical polar coordinates, $(\theta,\phi)$, the tensor spherical harmonics
are given explicitly by \cite{albert,zerilli},
\begin{equation}
   Y_{(lm)ab}^{\rm G}(\hatn)={N_l\over 2} \left( \begin{array}{cc}
   W_{(lm)}(\hatn) & X_{(lm)}(\hatn) \sin\theta\\
   \noalign{\vskip6pt}
   X_{(lm)}(\hatn)\sin\theta & -W_{(lm)}(\hatn)\sin^2\theta \\
   \end{array} \right)
\label{YGexplicit}
\end{equation}
and
\begin{equation}
   Y_{(lm)ab}^{\rm C}(\hatn)={N_l\over 2} \left( \begin{array}{cc}
   -X_{(lm)}(\hatn) & W_{(lm)}(\hatn) \sin\theta \\
   \noalign{\vskip6pt}
   W_{(lm)}(\hatn)\sin\theta & X_{(lm)}(\hatn)\sin^2\theta \\
   \end{array} \right),
\label{YCexplicit}
\end{equation}
where
\begin{equation}
     W_{(lm)}(\hatn) = \left( {\partial^2 \over \partial\theta^2} -
     \cot\theta {\partial \over \partial\theta} +
     {m^2\over\sin^2\theta}\right)
     Y_{(lm)}(\hatn) = 2 \left( {\partial^2 \over
     \partial\theta^2} - l(l+1) \right) Y_{(lm)}(\hatn),
\label{Wdefn}
\end{equation}
and
\begin{equation}
     X_{(lm)}(\hatn) = {2im \over \sin\theta}
     \left( {\partial \over \partial\theta} -
     \cot\theta \right) Y_{(lm)}(\hatn).
\label{Xdefn}
\end{equation}
(Note that this definition of $X_{(lm)}(\hatn)$ differs from
that in Ref. \cite{zerilli} by a factor of $\sin\theta$.)
The exchange symmetry $\{Q,U\}\leftrightarrow\{U,-Q\}$ as
G$\leftrightarrow$C indicates that $Y_{(lm)ab}^{\rm G}$ and
$Y_{(lm)ab}^{\rm C}$ represent polarizations rotated by
$45^\circ$.
By evaluating the derivatives, these functions can be written
\begin{equation}
     W_{(lm)}(\hatn) = -2
     \sqrt{ {2l+1 \over 4\pi} {(l-m)! \over
     (l+m)!} } \, G_{(lm)}^+(\cos\theta)\, e^{im\phi},
\label{WGplus}
\end{equation}
\begin{equation}
     i X_{(lm)}(\hatn) =  -2 
     \sqrt{ {2l+1 \over 4\pi} {(l-m)! \over
     (l+m)!} } \, G_{(lm)}^-(\cos\theta)\, e^{im\phi},
\label{XGminus}
\end{equation}
where the real functions $G_{(lm)}^\pm$ are defined by \cite{albert}
\begin{equation}
     G_{(lm)}^+(\cos\theta) \equiv - \left( {l-m^2 \over
     \sin^2\theta} + {1\over2} l(l-1) \right) P_l^m(\cos\theta)
     +(l+m) {\cos\theta \over \sin^2\theta}
     P_{l-1}^m(\cos\theta),
\label{Gplus}
\end{equation}
\begin{equation}
     G_{(lm)}^-(\cos\theta) \equiv {m\over\sin^2\theta} 
     \Bigl( (l-1) \cos\theta
     P_l^m(\cos\theta) - (l+m)P_{l-1}^m(x) \Bigr).
\label{Gminus}
\end{equation}
These expressions will be useful for the correlation functions
in Section~V, and for simulating maps and data analysis.

In linear theory,
scalar perturbations can produce only G-type polarization and not C-type
polarization.  On the other hand, tensor or vector metric perturbations
will produce a mixture of both types \cite{letter,sz96b,nextpaper}.
Heuristically,
this is because scalar perturbations have no handedness so they
cannot produce any ``curl'', whereas vector and tensor
perturbations {\it do} have a handedness and therefore can.
Observation of a non-zero primordial component of C-type polarization (a
nonzero $a_{(lm)}^{\rm C}$) in the CMB would
provide compelling evidence for significant contribution of either vector or
tensor perturbations at the time of last scattering.  

Given a polarization map of even a small part of the sky, one
could in principle test for vector or tensor contribution by
computing the combination of derivatives of the
polarization field given by ${\cal
P}^{ab}{}_{:bc}\epsilon^c{}_a$ which will be non-zero only for
C-type polarization.  Of course, taking derivatives of noisy data is
problematic.  We discuss more robust probes of this signal
below.

\section{Statistics of the Multipole Coefficients}

\subsection{Statistical Independence of the Coefficients}

We now have three sets of multipole moments, $a_{(lm)}^{\rm T}$,
$a_{(lm)}^{\rm G}$, and $a_{(lm)}^{\rm C}$,
which fully describe the
temperature/polarization map of the sky. 
Statistical isotropy implies that
\begin{eqnarray}
\VEV{a_{(lm)}^{{\rm T}\,*}a_{(l'm')}^{\rm T}}=
                   C^{\rm T}_l \delta_{ll'}\delta_{mm'},&\qquad\qquad&
\VEV{a_{(lm)}^{{\rm G}\,*}a_{(l'm')}^{\rm G}}= 
                                   C_l^{\rm G} \delta_{ll'}\delta_{mm'},    \cr
\VEV{a_{(lm)}^{{\rm C}\,*}a_{(l'm')}^{\rm C}}=
                   C_l^{\rm C} \delta_{ll'}\delta_{mm'},&\qquad\qquad&
\VEV{a_{(lm)}^{{\rm T}\,*}a_{(l'm')}^{\rm G}}=
                                   C_l^{\rm TG}\delta_{ll'}\delta_{mm'},    \cr
\VEV{a_{(lm)}^{{\rm T}\,*}a_{(l'm')}^{\rm C}}=
                   C_l^{\rm TC}\delta_{ll'}\delta_{mm'},&\qquad\qquad&
\VEV{a_{(lm)}^{{\rm G}\,*}a_{(l'm')}^{\rm C}}=
                                   C_l^{\rm GC}\delta_{ll'}\delta_{mm'},
\label{cldefs}
\end{eqnarray}
where the angle brackets are an average over all realizations.
For Gaussian theories, the statistical properties of a
temperature/polarization map are specified fully by these six
sets of multipole moments.
In fact, the scalar spherical harmonics $Y_{(lm)}$ and the G
tensor harmonics $Y_{(lm)ab}^{\rm G}$ have parity $(-1)^l$, but the C
harmonics $Y_{(lm)ab}^{\rm C}$ have parity $(-1)^{l+1}$.
Therefore, symmetry under parity transformations requires that
$C_l^{\rm TC}=C_l^{\rm GC}=0$, which will 
also be demonstrated explicitly in the following
Section.  Measurement of nonzero cosmological
values for these moments would
be quite extraordinary, demonstrating a handedness
to primordial perturbations. In practice, these two
sets of moments can be used to monitor foreground
emission. Furthermore, as mentioned above and demonstrated
explicitly in Sec.~IV, $C_l^{\rm C}=0$ for scalar
metric perturbations \cite{letter,sz96b}. 
At small angular scales where the
contribution from tensor and vector perturbations is expected
to be negligible, $C_l^{\rm C}$ can also be pressed into
duty as a foreground monitor.
Exact expressions for these multipole moments
in terms of the photon brightnesses
usually calculated by early-Universe Boltzmann codes are 
derived below.

\subsection{Map Simulation}

For the case of Gaussian statistics, realizations of
temperature/polarizations maps are easy to generate
using standard techniques.  Since the only
cross-correlation between mode coefficients, given by $C_l^{\rm TG}$,
correlates only  $a_{(lm)}^{\rm T}$ and $a_{(lm)}^{\rm G}$ with the same $l$
and $m$, the total correlation matrix is block diagonal with the largest blocks
being only 2$\times$2 matrices.  In particular, set
\begin{eqnarray}
a_{(lm)}^{\rm T}&=&\zeta_1\left(C_l^{\rm T}\right)^{1/2},          \cr
a_{(lm)}^{\rm G}&=&\zeta_1
                             {C_l^{\rm TG}\over\left(C_l^{\rm T}\right)^{1/2}}
                  +\zeta_2
   \left( C_l^{\rm G}-
         {\left(C_l^{\rm TG}\right)^2\over C_l^{\rm T}}\right)^{1/2},       \cr
a_{(lm)}^{\rm C}&=&\zeta_3\left(C_l^{\rm C}\right)^{1/2},
\end{eqnarray}
where for each value of $l$ and $m>0$ choose three complex numbers,
$(\zeta_1,\zeta_2,\zeta_3)$, drawn from a Gaussian distribution with unit
variance, i.e., both $\sqrt{2}\,{\rm Re}(\zeta_i)$ and
$\sqrt{2}\,{\rm Im}(\zeta_i)$ are drawn from a normal distribution.  For $m=0$
the same equations hold but the $\zeta_i$ should be real and normally
distributed; for $m<0$ the coefficients are given by Eq.~(\ref{almsymmetry}).  
Note 
that in all cases $C_l^{\rm G}C_l^{\rm T}\ge\left(C_l^{\rm TG}\right)^2$.  This
set of coefficients can be combined with Eqs.~(\ref{whatPis}),
(\ref{Pexpansion}), (\ref{Xdefn}), and (\ref{Wdefn}) to obtain the explicit
expressions 
\begin{eqnarray}
     Q(\hatn)&=&2{\cal P}_{\theta\theta}(\hatn)
      = T_0 \sum_{l=2}^\infty\sum_{m=-l}^l N_l 
        [a_{(lm)}^{\rm G} W_{(lm)}(\hatn)-a_{(lm)}^{\rm C} X_{(lm)}(\hatn)] \cr
     U(\hatn)&=&-2\csc\theta{\cal P}_{\theta\phi}(\hatn)
      =-T_0 \sum_{l=2}^\infty\sum_{m=-l}^l N_l 
        [a_{(lm)}^{\rm G} X_{(lm)}(\hatn)+a_{(lm)}^{\rm C} W_{(lm)}(\hatn)]
\label{QUexpansion}
\end{eqnarray}
with $W_{(lm)}$ and $X_{(lm)}$ given by Eqs.~(\ref{WGplus}) 
and (\ref{XGminus}).
Note that polarization maps are traditionally plotted as headless vectors with
amplitude $(Q^2+U^2)^{1/2}$ and orientation angle $(1/2)\tan^{-1}(U/Q)$.

\subsection{Estimators}

One of the the main uses 
of a temperature/polarization map will be to determine
the multipole moments with the best possible accuracy.  From a
full-sky CMB temperature map, we can construct the following
rotationally-invariant estimators (denoted by a $\widehat{\ \ }$)
for the multipole coefficients: 
\begin{eqnarray}
\widehat{C_l^{\rm T}}=\sum_{m=-l}^l{\left|a^{\rm T}_{(lm)}\right|^2\over2l+1},
&\qquad\qquad&
\widehat{C_l^{\rm G}}=\sum_{m=-l}^l{\left|a^{\rm G}_{(lm)}\right|^2\over2l+1},
\cr
\widehat{C_l^{\rm C}}=\sum_{m=-l}^l{\left|a_{(lm)}^{\rm C}\right|^2\over2l+1},
&\qquad\qquad&
\widehat{C_l^{\rm TG}}=\sum_{m=-l}^l{a_{(lm)}^{\rm T\,*}\,
                                     a_{(lm)}^{\rm G}\over2l+1}.
\label{estimators}
\end{eqnarray}
Note that Eq.~(\ref{almsymmetry}) guarantees that the $\widehat{C_l^{\rm TG}}$
will be real.

When averaged over the sky (denoted by an overbar), the mean square
temperature anisotropy after subtracting the dipole is
\begin{equation}
    \overline{\left( { \Delta T \over T_0} \right)^2} = \sum_{l=2}^\infty
    {2l+1 \over 4\pi} \widehat{C_l^{\rm T}},
\label{meansquareT}
\end{equation}
and the mean square polarization is
\begin{equation}
\overline{{\cal P}^2} \equiv \overline{Q^2+U^2} 
            =2\overline{{\cal P}^{ab} {\cal P}_{ab}}
            =\overline{{\cal P_{\rm G}}^2}+\overline{{\cal P_{\rm C}}^2},
\nonumber
\end{equation}
\begin{equation}
\overline{{\cal P_{\rm G}}^2}
         \equiv T_0^2 \sum_{l=2}^\infty{2l+1\over 8 \pi}\widehat{C_l^{\rm G}} 
\qquad\qquad
\overline{{\cal P_{\rm C}}^2}
         \equiv T_0^2 \sum_{l=2}^\infty{2l+1\over 8 \pi}\widehat{C_l^{\rm C}}.
\label{meansquareP}
\end{equation}
Even if no single $\widehat{C_l^{\rm C}}$ or
$\widehat{C_l^{\rm G}}$ gives a significant signal,
combining different $l$'s as in $\overline{{\cal P_{\rm G}}^2}$ or
$\overline{{\cal P_{\rm C}}^2}$ can give a statistically significant
signal.

\subsection{Cosmic and Pixel-Noise Variance}

The averages in Eqs. (\ref{cldefs}) are over an ensemble of universes drawn
{}from a theoretically defined statistical distribution, or
assuming ergodicity, a spatial average over 
all observer positions in the Universe.  However, we can only
observe a single realization of the ensemble from a single
location.  Therefore, even if we had an ideal (full-sky
coverage, no foreground contamination, infinite angular
resolution, and no instrumental noise) experiment, the accuracy
with which the estimators in
Eqs.~(\ref{estimators}) could recover
the multipole moments would be limited by a sample variance
known as ``cosmic variance.''  Furthermore, a realistic
experiment may have limited sky coverage and angular resolution
and some instrumental noise.  In this Section, we calculate the
cosmic variance with which the multipole moments can be
recovered.  
We also calculate the variance due to finite sky
coverage, angular resolution, and instrumental noise
in an idealized experiment.  To do so,
we adopt a simplified model in which we assume a pixelized map in
which the noise in each pixel is independent and Gaussian
distributed {\it after} foregrounds have been successfully
subtracted.  In many respects, our derivation follows that in
Ref.~\cite{knox}, and our results agree with those in
Ref.~\cite{sz96c}.

We must first determine the contribution of pixel noise to each
multipole moment, and we begin with the temperature moments.
Consider a temperature map of the full sky, $T^{\rm
map}(\hatn)$, which is pixelized with $N_{\rm pix}$ pixels.
If we assume that each pixel subtends the same area on the sky then
we can construct multipole coefficients of the temperature map using
\begin{equation}
     d_{(lm)}^{{\rm T}} = \int d\hatn\,  \left({T^{\rm
     map}(\hatn) \over T_0} \right)\,
     Y_{(lm)}(\hatn) \simeq {1\over T_0} \sum_{j=1}^{N_{\rm
     pix}} {4\pi \over N_{\rm pix}}\,  T^{\rm map}_j\,
     Y_{(lm)}(\hatnj),
\end{equation}
where $T^{\rm map}_j$ is the measured temperature
perturbation in pixel $j$, and $\hatnj$ is its direction.  
The difference between $d_{(lm)}$ and $a_{(lm)}$ is that the former includes
the effects of finite beam size and detector noise, i.e. $d_{(lm)}$ is 
measured.
The extent to which the approximate equality fails is the
pixelization noise, which is small on angular scales much larger
than the pixels.
In many of the proposed experiments, which oversample the sky when compared to
their beam, there will be no loss of information by ignoring scales close to
the pixel scale.  The observed temperature is due to a
cosmological signal and a pixel noise, $T^{\rm map}_j = T_j
+T^\noise_j$.  If we assume that each pixel has the same rms noise, and
that the noise in each pixel is uncorrelated with that in any other pixel, and
is uncorrelated with the cosmological signal, i.e. 
$\VEV{T^\noise_i T^\noise_j} 
=T_0^2\left(\sigma_{\rm pix}^{\rm T}\right)^2\delta_{ij}$ and
$\VEV{T_i\,T^\noise_j}=0$, then
\begin{eqnarray}
     \VEV{d_{(lm)}^{\rm T} d_{(l'm')}^{{\rm T}\,*}} & = &
     \VEV{a_{(lm)}^{\rm T} a_{(l'm')}^{{\rm T}\,*}}+
     \VEV{a_{(lm)}^{{\rm T},\noise}
     \left(a_{(l'm')}^{{\rm T},\noise}\right)^*} \nonumber \\
     &= & |W_l^{\rm b}|^2 C_l^{\rm T}  \delta_{ll'}\delta_{mm'} 
    +\VEV{a_{(lm)}^{{\rm T},\noise}\left(a_{(l'm')}^{{\rm T},\noise}\right)^*},
\label{mapmoments}
\end{eqnarray}
where we have written the expectation value of the cosmological signal
in terms of that predicted by theory, $C_l^{\rm T}$, multiplied
by $|W_l^{\rm b}|^2$
which accounts for beam smearing.  Typically the beam is approximately Gaussian
in shape, corresponding to the window function 
$W_l^{\rm b}\approx\exp(-l^2\sigma_{\rm b}^2/2)$, where
$\sigma_{\rm b}=\theta_{\rm fwhm}/\sqrt{8\ln2}\,
=0.00742\,(\theta_{\rm fwhm}/1^\circ)$ and $\theta_{\rm fwhm}$ gives the
full-width at half maximum.

The second term in Eq.~(\ref{mapmoments}) is
\begin{eqnarray}
     \VEV{ a_{(lm)}^{{\rm T},\noise} \left(a_{(l'm')}^{{\rm T},\noise}
     \right)^*} & = & {1\over T_0^2}\sum_{i=1}^{N_\pix}\sum_{j=1}^{N_\pix}
     \left( {4\pi \over N_{\rm pix}}\right)^2 
     \VEV{T^\noise_i T^\noise_j}\,Y_{(lm)}(\hatni)\,Y^*_{(l'm')}(\hatnj)
     \nonumber \\
	&= &\left(\sigma_{\rm pix}^{\rm T}\right)^2 \left(
	\sum_{i=1}^{N_\pix} {4\pi \over N_\pix}\, 
     Y_{(lm)}(\hatni) \,  Y_{(l'm')}(\hatnj) \right) {4\pi \over
     N_\pix} \nonumber \\
	& =& {4\pi \left(\sigma_{\rm pix}^{\rm T}\right)^2 \over
	N_\pix }\, \delta_{ll'}\, 
     \delta_{mm'}.
\end{eqnarray}
Therefore, the moments measured by the map are distributed with
a variance \cite{knox}
\begin{equation}
     \VEV{d_{(lm)}^{\rm T} d_{(l'm')}^{{\rm T}\,*}} =
     \left(C_l |W_l^{\rm b}|^2 + {4\pi \left(\sigma_{\rm pix}^{\rm T}\right)^2
     \over N_\pix} \right) \delta_{ll'} \delta_{mm'}.
\label{sigmalmap}
\end{equation}

Now we move on to the noise contribution to the polarization moments.   We will
assume the instrumental noise in the polarization measurements is isotropic,
the same for all pixels, uncorrelated with the noise in the anisotropy, which
in terms of the Stokes parameters requires
\begin{equation}
      \VEV{Q_i^\noise Q_j^\noise}=\VEV{U_i^\noise U_j^\noise} 
                                 =T_0^2 (\sigma_\pix^{\rm P})^2 \delta_{ij},
\qquad\VEV{Q_i^\noise U_j^\noise}=
      \VEV{Q_i^\noise T_j^\noise}=\VEV{U_i^\noise T_j^\noise}=0 .
\end{equation}
We denote the polarization tensor describing the noise in pixel $i$ by
${\cal P}^\noise_{ab}(\hatn_i)$. The previous equations are equivalent to
the coordinate independent equation (see ref~\cite{albert})
\begin{equation}
\VEV{{\cal P}^\noise_{ab}(\hatn_i)\,{\cal P}^\noise_{cd}(\hatn_j)}
={1\over4}\,T_0^2(\sigma_\pix^{\rm P})^2
(g_{ac}g_{bd}-\epsilon_{ac}\epsilon_{bd})\,\delta_{ij} \qquad
\VEV{{\cal P}^\noise_{ab}(\hatn_i)\,T^\noise(\hatn_j)}=0 ,
\label{PolarizationWhiteNoise}
\end{equation}
The mode coefficients for the noise, defined as in
Eq.~(\ref{defmoments}), will have a correlation matrix
\begin{equation}
\VEV{a_{(lm)}^{X,\noise\,*}a_{(l'm')}^{X',\noise\,*}}
=\left({4\pi\over N_\pix  T_0}\right)^2\,\sum_{i=1}^{N_\pix}\sum_{j=1}^{N_\pix}
            Y_{(l m )}^{X \,ab\,*}(\hatn_i)\,Y_{(l'm')}^{X'\,cd   }(\hatn_j)\,
            \VEV{{\cal P}_{ab}(\hatn_i){\cal P}_{cd}(\hatn_j)}    \qquad
                                                          X,X'\in\{ {\rm G,C}\}
\end{equation}
so using
Eqs.~(\ref{PolarizationWhiteNoise}), (\ref{norms}), 
and (\ref{ShotNoiseIdentity})
we find
\begin{equation}
\VEV{a_{(lm)}^{X,\noise\,*}a_{(l'm')}^{X',\noise\,*}}
={1\over4}\left({4\pi\sigma_\pix^{\rm P}\over N_\pix}\right)^2\,
   \sum_{i=1}^{N_\pix}2\,Y_{(l m )}^{X \,ab\, *}(\hatn_i)\,
                         Y_{(l'm')ab}^{X'}      (\hatn_i)
={2\pi\over N_\pix}\,(\sigma_\pix^{\rm P})^2\,
                     \delta_{ll'}\delta_{mm'}\delta_{XX'}
\end{equation}
and of course
\begin{equation}
\VEV{a_{(lm)}^{{\rm G},\noise\,*}a_{(l'm')}^{{\rm T},\noise}}=
\VEV{a_{(lm)}^{{\rm C},\noise\,*}a_{(l'm')}^{{\rm T},\noise}}=0 .
\end{equation}
Thus instrumental noise contributes equally to the measured G- and
C- polarization components but introduces no cross-correlation between them.

Collecting the results, if $d_{(lm)}^{\rm X}$ (for
 ${\rm X}=\{{\rm T},{\rm G},{\rm C}\}$) are the multipole coefficients for
the map (signal plus noise), their variances will be
\begin{equation}
     \VEV{ d_{(lm)}^{{\rm X}\,*} d_{(l'm')}^{{\rm X}'}}
     =(|W_l^{\rm b}|^2 C_l^{{\rm XX}'}+w_{{\rm XX}'}^{-1})
                                                      \delta_{ll'} \delta_{mm'}
     \equiv D_l^{{\rm XX}'} \delta_{ll'} \delta_{mm'},
\label{mapautovariances}
\end{equation}
where 
\begin{equation}
     w_{\rm T}^{-1}\equiv{4\pi\left(\sigma_\pix^{\rm T}\right)^2\over N_\pix},
\qquad\qquad
     w_{\rm GG}^{-1}=w_{\rm CC}^{-1}\equiv w_{\rm P}^{-1}
                   \equiv{2\pi\left(\sigma_\pix^{\rm P}\right)^2 \over N_\pix},
\label{winverse}
\end{equation}
and $w_{{\rm XX}'}^{-1}=0$ for ${\rm X}\neq {\rm X}'$.
The quantities $w^{-1}$ are inverse statistical weights per unit
solid angle, a measure of experimental sensitivity independent
of pixel size \cite{knox}.
(Note that our $w_{\rm P}^{-1}$ differs by a factor of two from
that in Ref. \cite{sz96c} which is consistent with 
our $C_l^{\rm G}$ and $C_l^{\rm C}$ differing from their E and B
moments by a factor of two.)

Estimators for the multipole moments CMB power spectra,
$C_l^{{\rm XX}'}$, are
\begin{equation}
     \widehat{C_l^{{\rm XX}'}} = \left( \widehat{D_l^{{\rm XX}'}} -
     w_{{\rm XX}'}^{-1} \right) |W_l^{\rm b}|^{-2},
\end{equation}
where
\begin{equation}
     \widehat{D_l^{{\rm XX}'}}= \sum_{m=-l}^l {d_{(lm)}^{{\rm
     X}\,*} d_{(lm)}^{\rm X'} \over 2l+1}.
\end{equation}

Since the estimators for each $D_l^{{\rm XX}'}$ (and therefore for
$C_l^{{\rm XX}'}$) are constructed from only $2l+1$ multipole
coefficients, each $C_l^{{\rm XX}'}$ can be
recovered only with a finite sampling variance, known as
cosmic variance.  In addition, the six different sets of measured
moments are constructed from the three sets of $d_{(lm)}$
coefficients, leading to some covariance between the
moments.  These variances can be described with a
($6\times 6$) covariance matrix,
\begin{eqnarray}
     \Xi_{{\rm AA}'} & \equiv & \VEV{ \left( \widehat{C_l^{\rm
     A}} - C_l^{\rm A} \right)
     \left( \widehat{C_l^{{\rm A}'}} - C_l^{{\rm A}'} \right) } \nonumber \\
     & = & \VEV{ \widehat{C_l^{\rm A}} \widehat{C_l^{{\rm A}'}} } -
     C_l^{\rm A} C_l^{{\rm A}'}
     \nonumber \\
     & = & \left( \VEV{ \widehat{D_l^{\rm A}} \widehat{D_l^{{\rm A}'}} } -
     D_l^{\rm A} D_l^{{\rm A}'} \right) |W_l^{\rm b}|^{-4}
\label{Xidefn}
\end{eqnarray}
for A$={\rm XX}'$.

We now calculate the entries of this matrix.  
Recall that if $x_i$ are Gaussian random variables
with variances $\VEV{x_i x_j} = \sigma_{ij}^2$, then $\VEV{x_i^2
x_j^2} = \sigma_{ii}^2 \sigma_{jj}^2 + 2 \sigma_{ij}^2$, and
$\VEV{x_i^3 x_j} = 3 \sigma_{ii}^2 \sigma_{ij}^2$.
For X$=$\{T,G,C\},
\begin{eqnarray}
     \VEV{ \widehat{D_l^{\rm XX}} \widehat{ D_l^{{\rm X}'{\rm
     X}'}} }  & = & \sum_{mm'} { \VEV{ |d_{(lm)}^{\rm X} |^2
     |d_{(lm')}^{{\rm X}'} |^2} \over (2l+1)^2 } \nonumber \\
     &= &\sum_{mm'} {1\over(2l+1)^2}
     \left[D_l^{\rm XX} D_l^{{\rm X}'{\rm X}'} (1-\delta_{mm'})
     + \left( D_l^{\rm XX} D_l^{{\rm X}'{\rm X}'} + 2 \left(
     D_l^{{\rm XX}'} \right)^2 \right) \delta_{mm'} \right]
     \nonumber \\
     & = & D_l^{\rm XX} D_l^{{\rm XX}'} + {2 \over (2l+1)^2}
     \left( D_l^{{\rm XX}'} \right)^2.
\end{eqnarray}
The diagonal elements for TT, GG, and CC are thus
\begin{equation}
     \Xi_{\rm XX,XX} = {2 \over 2l+1} \left( C_l^{\rm XX} +
     w_{\rm X}^{-1} |W_l^{\rm b}|^{-2} \right)^2,
\label{diagonalvariances}
\end{equation}
and the off-diagonal elements are
\begin{eqnarray}
     \Xi_{\rm TT,GG} &=& {2 \over 2l+1} \left( C_l^{\rm TG}\right)^2,
     \nonumber \\
     \Xi_{\rm TT,CC} &=& 0, \nonumber \\
     \Xi_{\rm GG,CC} &=& 0.
\end{eqnarray}

For the diagonal TG component,
\begin{eqnarray}
     \VEV{ \widehat{D_l^{\rm TG}} \widehat{D_l^{\rm TG}}} & = &
     \sum_{mm'} { \VEV{ d_{(lm)}^{{\rm T}\,*}
     d_{(lm)}^{\rm G} d_{(lm')}^{{\rm T}\,*}
     d_{(lm')}^{\rm G} } \over (2l+1)^2} \nonumber \\
     &=& {1\over (2l+1)^2}
     \sum_{mm'}\left[  \VEV{|d_{(lm)}^{\rm T}|^2 |d_{(lm)}^{\rm
     G}|^2} \delta_{mm'} + \VEV{ d_{(lm)}^{{\rm T}\,*}
     d_{(lm)}^{\rm G} d_{(lm')}^{{\rm T}}
     d_{(lm')}^{{\rm G}\,*} } (1-\delta_{mm'}) \right]
     \nonumber \\
     & =& \sum_m {D_l^{\rm TT} D_l^{\rm GG} + 2 \left( D_l^{\rm
     TG} \right)^2 \over (2l+1)^2} + \sum_{mm'}
     {(1-\delta_{mm'}) \left( D_l^{\rm TG}\right)^2 \over
     (2l+1)^2} \nonumber \\
     &=& {1\over 2l+1} \left[ \left(D_l^{\rm TG} \right)^2 +
     D_l^{\rm TT} D_l^{\rm TG} \right] + \left(D_l^{\rm TG}
     \right)^2.
\end{eqnarray}
Therefore,
\begin{equation}
     \Xi_{\rm TG,TG} = {1\over 2l+1} \left[ \left(C_l^{\rm TG}\right)^2
      +\left( C_l^{\rm T} + w_{\rm T}^{-1} |W_l^{\rm b}|^{-2} \right)
       \left( C_l^{\rm G} + w_{\rm P}^{-1} |W_l^{\rm b}|^{-2} \right) \right].
\label{diagonalTGvariance}
\end{equation}
The diagonal covariance-matrix elements for TC and TG are
obtained similarly and are nonzero even though $C_l^{\rm
TC}=C_l^{\rm GC}=0$.  Given a map, it should be checked for
consistency that there is no statistically significant parity
violation.

Finally, for the off-diagonal TG-TT component,
\begin{eqnarray}
     \VEV{ \widehat{D_l^{\rm TG}} \widehat{D_l^{\rm TT}}} & =&
     \sum_{mm'} {1\over (2l+1)^2} 
     \left[\VEV{ |d_{(lm)}^{\rm T}|^3 d_{(lm)}^{\rm G}}
     \delta_{mm'} + \VEV{ |d_{(lm)}^{\rm T}|^2 (d_{(lm')}^{\rm
     T})^* d_{(lm')}^{\rm G}} (1-\delta_{mm'})\right]
     \nonumber \\
     &=& \left({2 \over 2l+1} +1 \right) D_l^{\rm TT} D_l^{\rm
     TG}.
\end{eqnarray}
Therefore,
\begin{equation}
     \Xi_{\rm TT,TG} = {2 \over 2l+1} \left[C_l^{\rm TG} \left(
     C_l^{\rm T} + w_{\rm T}^{-1} |W_l^{\rm b}|^{-2} \right)
     \right],
\end{equation}
and similarly
\begin{equation}
     \Xi_{\rm GG,TG} = {2 \over 2l+1} \left[C_l^{\rm TG} \left(
     C_l^{\rm G} + w_{\rm G}^{-1} |W_l^{\rm b}|^{-2} \right)
     \right],
\label{offdiagonalvariances}
\end{equation}
and this completes the calculation of all nonzero elements of
the covariance matrix.

To determine the precision with which a
temperature/polarization map can recover some cosmological
parameters, e.g.,  ${\bf s} =\{ \Omega_0, \Lambda, h,
\Omega_b,... \}$, we can evaluate the curvature, or
Fisher information, matrix \cite{fisher}, which can be generalized from
that in Ref.~\cite{parameters} to
\begin{equation}
     \alpha_{ij} = \sum_l \sum_{{\rm A,A'}} { \partial C_l^{\rm
     A} \over \partial s_i} \left[\Xi^{-1}\right]_{AA'}  { \partial C_l^{\rm
     A'} \over \partial s_j},
\end{equation}
for A$=$TT,GG,CC,TG where $[\Xi^{-1}]_{AA'}$ are elements of the
inverse of $\Xi$.  The standard error with which a given
parameter $s_i$ can be recovered after marginalizing over all
other parameters is given by the square root of the diagonal
element $i$ of $[\alpha]^{-1}$, assuming a linear dependence
of $C_l^X$ on all of the parameters.

The $w_{\rm X}^{-1}$ factors on the right-hand side of
Eqs. (\ref{diagonalvariances}), (\ref{diagonalTGvariance}), and
(\ref{offdiagonalvariances}) are those due to 
instrumental noise.  However, note that even in an ideal
experiment with $\sigma_\pix=0$, the right-hand sides would still
be nonzero, and this is the cosmic variance.
Eqs. (\ref{diagonalvariances}), (\ref{diagonalTGvariance}), and
(\ref{offdiagonalvariances}) are valid for a map
with full-sky coverage.  Realistically, however, only a fraction
$f_{\rm sky}$ of the sky will be surveyed, or if the entire sky
is surveyed, only a fraction will be used in the analysis.
Therefore, the the accuracy with which cosmological parameters
can be recovered will be degraded accordingly.  Strictly
speaking, harmonic analysis on a cut sky will have to be
performed, and this will introduce correlations in the errors of
multipole moments of different $l$'s.  However, if the entire
sky is surveyed, but only a fraction of the sky is used for the
analysis (e.g., if the Galactic plane has been cut out), then
the effect of partial sky coverage can be approximated by
multiplying the curvature matrix by $f_{\rm sky}$ (which will
increase the standard errors in cosmological parameters by
$f_{\rm sky}^{-1/2}$).  If only a fraction of the sky is
surveyed, the curvature matrix should still be multiplied by
$f_{\rm sky}$, but note that $N_\pix$ is the
number of pixels actually in the map.

\subsection{Pixel Noise for a Polarization Map}

How are the temperature and polarization pixel noises 
related?  
If the two linear polarization states
are always given equal integration times, the total number
of photons available for the temperature measurement will be
twice the number available for either polarization measurement.
Therefore,
\begin{equation}
     \left(\sigma_\pix^{\rm T} \right)^2 =
     {1\over2}\left(\sigma_\pix^{\rm P} \right)^2.
\end{equation}

However, a crucial difference in overall sensitivity exists
between the two current receiver technologies. Coherent
receivers (i.e. HEMT amplifiers) measure the incoming
electric field, while incoherent receivers (i.e. bolometers)
measure only the incoming total power.
In the former case, the signal can be split into two orthogonal
linear polarizations and the phase information can be retained
throughout the entire signal path. If properly designed,
such a system of receivers can measure the linear
polarizations without compromising the temperature measurement,
so the temperature sensitivity will be determined solely
by the amplifier characteristics \cite{mapsatellite}.
This is the design strategy for the MAP satellite.

On the other hand, bolometers measure only the instantaneous
total power received and do not retain any phase information;
a filter must be placed in front of the detector
for each linear polarization state, 
discarding half of the incoming photons. 
Thus the temperature sensitivity for a polarized
bolometer measurement is
only half of that for an unpolarized measurement,
which collects twice as many photons in the same amount
of integration time. 
A compensating factor is that bolometers offer much greater raw sensitivity
than HEMT amplifiers. An important question facing future
bolometer experiments is whether to sacrifice half of the
temperature sensitivity for polarization information \cite{paramnext}. If
the goal of an experiment is to measure angular power spectra and
the temperature measurements are dominated by cosmic variance,
then polarizing the measurement is obviously advantageous.
The answer is as yet unclear in cases where cosmic variance
is not the controlling factor in the temperature measurement.

\section{Calculation of the Moments}

In this Section, we calculate the set of multipole
moments defined by Eqs.~(\ref{cldefs}), for scalar
and tensor metric perturbations.
Vector metric perturbations make a negligible contribution for
inflationary theories, although they are generic in defect models;
they will be covered elsewhere.
The intensity and linear-polarization state of the CMB in
any given direction is specified by three quantities (the
temperature $T$ and the Stokes parameters
$Q$, and $U$), giving six possible sets of multipole moments,
$C_l^{\rm T}$, $C_l^{\rm G}$, $C_l^{\rm C}$, $C_l^{\rm TG}$, $C_l^{\rm TC}$, 
and
$C_l^{\rm GC}$, but as argued above, parity demands that
$C_l^{\rm TC}=C_l^{\rm GC}=0$. 

One way to calculate the moments is to rewrite the
radiative transfer equations in terms of tensor
harmonics \cite{nextpaper}.  
The contribution of each Fourier mode to each
multipole moment is then obtained by evolving numerically the coupled
Einstein and Boltzmann equations for the multipole coefficients
$a_{(lm)}^{\rm X}$.  Integrating over
all Fourier modes then gives the multipole moments.  This approach
has the advantage of being similar in form to the usual
moment hierarchy formulation of the problem, while keeping
the independent modes separated throughout the calculation,
giving simple power spectrum expressions.
A second approach offering computational advantages has been
presented in Ref.~\cite{sz96c}, which uses the Stokes parameter
evolution equations to write an integral equation solution for
the multipole moments.

Another possibility, which makes contact with previous work on
CMB temperature anisotropies and polarization, is to express the
multipole moments in terms of the usual perturbations 
to the photon brightness and polarization, $\Delta_{Il}$
and $\Delta_{Ql}$ (see Ref. \cite{annals} for definitions and
descriptions), obtained from most current numerical
calculations \cite{codes}, and this is the
approach adopted here.  In this Section, we calculate the
contributions to these moments from scalar and tensor
perturbations.  
With these results, it is straightforward to
modify existing numerical codes to obtain all of the multipole
moments.

\subsection{Scalar Metric Perturbations}

The simplest calculation of the multipole moments
in terms of the photon brightnesses uses the fact
that, due to statistical isotropy, the contribution of a given
$\veck$ mode to the moments $C_l$ depends on its magnitude only
and not its direction.  Therefore, we will consider the
contribution of a single $\veck$ mode, 
with the coordinate system always chosen with $\hatz$ in the
$\bf k$ direction, and
then integrate over all $\veck$ at the end.
All temperatures, Stokes parameters, polarization tensors,
and expansion coefficients are functions of $\bf k$, although
we sometimes drop explicitly reference to $\bf k$ for notational 
simplicity. Functions of the real-space coordinate $\bf x$ do
not appear in this paper.

\subsubsection{Temperature moments}

We begin with the familiar temperature moments.  The temperature
anisotropy induced in the direction $\hatn$ on the sky by a
single $\veck$ scalar mode is given by Eq. (7.1) in
Ref.~\cite{annals},
\begin{equation}
{T({\bf k},\hatn)\over T_0}
={1\over4}\sum_{j=1}^\infty(2j+1)P_j(\hatk\cdot\hatn)\,\Delta_{Ij}^{\rm
s}(k)
=1+\sum_{l=2}^\infty\sum_{m=-l}^l 
a_{(lm)}^{{\rm T,\,s}}({\bf k})Y_{(lm)}(\hatn).
\end{equation}
The superscripts T and s indicate that we are dealing here
with temperature moments from scalar perturbations.
The $\Delta_{Il}^{\rm s}(k)$ are Legendre coefficients of the photon
intensity distribution function for scalar metric perturbations
\cite{annals}.  The expansion coefficients are given by
the inverse transform, Eq. (\ref{temperaturemoments}), 
\begin{eqnarray}
a_{(lm)}^{{\rm T,s}}(\veck)&=&{1\over4}\sum_{j=0}^\infty(2j+1)\Delta_{Ij}(k)
               \int d\hatn\, P_j(\hatk\cdot\hatn) Y_{(lm)}^*(\hatn)\nonumber \\
          &=&\pi\Delta_{Il}^{\rm s}(k)Y_{(lm)}^*(\hatk)
	  \nonumber \\ 
          &=&{1\over2}\sqrt{(2l+1)\pi}\,\Delta_{Il}^{\rm s}(k)\delta_{m0},
\label{scalarTalm}
\end{eqnarray}
where we have used Eq.~(\ref{PYconvolution}) and taking $\hatk=\hatz$ in the
final line.  The contribution to $C_l^{{\rm T,s}}$ from this mode is then
\begin{equation}
     C_l^{{\rm T,s}}(\veck)={1\over2l+1}\sum_{m=-l}^l
     |a_{(lm)}^{\rm T,s}(\veck)|^2
                           ={\pi\over4}|\Delta_{Il}^{\rm s}(k)|^2,
\end{equation}
The total multipole moment from scalar modes, $C_l^{\rm T,scalar}$, is
given by integrating over all $\veck$:
\begin{equation}
     C_l^{\rm T,scalar} = \int {d^3\veck \over (2\pi)^3} \,
     C_l^{{\rm T,s}}(\veck) =
     {1 \over 8\pi}\, \int \, k^2\,dk\, |\Delta_{Il}^{\rm
     s}(k)|^2,
\label{ClTs}
\end{equation}
which is the usual result.

\subsubsection{Polarization moments}

Now we move on to the polarization moments produced by scalar perturbations.
First we describe the ``standard'' representation of polarization, which is
what has usually been computed by Boltzmann codes.  In the spherical polar
coordinates, $(\theta,\phi)$, the Stokes parameters induced in direction
$\hatn$ on the sky by a single $\veck$ scalar mode in the ${\bf \hat z}$
direction can be obtained {}from Eq. (7.4) in Ref.~\cite{annals} (the $\xi'$
there can be chosen zero for $\hatk=\hatz$) and are
\begin{equation}
{Q({\bf k},\hatn)\over T_0}
={1\over4}\sum_{j=0}^\infty(2j+1)P_j(\hatk\cdot\hatn)\Delta_{Qj}^{\rm s}(k) 
\qquad U({\bf k},\hatn)=0,
\label{QUscalar}
\end{equation}
where the $\Delta_{Ql}^{\rm s}(k)$ are Legendre coefficients of
the photon polarization distribution function for scalar metric
perturbations.
The polarization tensor at a point $\hatn$ on the sky
induced by this scalar mode is thus
\begin{eqnarray}
     {{\cal P}_{ab}({\bf k},\hatn)\over T_0} 
     & = & {1\over8} \sum_{j=0}^\infty (2j+1)
     P_j(\hatk\cdot\hatn) \Delta_{Qj}^{\rm s}(k) \left(
     \begin{array}{cc}
     \vphantom{1\over 2}1 & 0 \\
     0 & -\sin^2\theta \\ 
     \end{array} \right) \nonumber \\
     & = & {1\over8} \sum_{j=0}^\infty (2j+1)
     \Delta_{Qj}^{\rm s}(k) M_{(j)\,ab}(\hatk,\hatn),
\label{Pscalarpattern}
\end{eqnarray}
where the second line defines the tensor $M_{(j)\,ab}$; i.e., 
the tensor $M_{(j)}^{ab}$ (with raised indices)
takes the form
\begin{equation}
     M^{ab}_{(j)}(\hatk={\bf\hat z},\hatn) = P_j(\cos\theta) \left(
     \begin{array}{cc}
     \vphantom{1\over 2}1 & 0 \\
     0 & {-\csc^2\theta} \\
     \end{array} \right).
\label{Msimple}
\end{equation}
This $j$-expansion is not an expansion in tensor spherical harmonics and, as we
shall see, the harmonic content of the $M^{ab}_{(j)}$ tensor, while peaked
around $l=j$, has significant contributions from $l$ far from $j$.

	We now proceed to re-express the above representation of 
polarization in
terms of tensor spherical harmonics.  The G multipole coefficients of the
pattern in Eq. (\ref{Pscalarpattern}), given by Eq. (\ref{Gmomentseasy}), are
\begin{eqnarray}
     a_{(lm)}^{{\rm G,s}}(\veck)
     & = & N_l \int d\hatn\, Y_{(lm)}^*(\hatn)
     {\cal P}^{ab}{}_{:ab}({\bf k},\hatn) \nonumber \\
     & = & {N_l \over 8} \sum_{j=0}^\infty (2j+1)
     \Delta_{Qj}^{\rm s}(k) 
     \int d\hatn Y_{(lm)}^*(\hatn) 
     M_{(j)}^{ab}{}_{:ab}(\hatn) .
\label{almE}
\end{eqnarray}
We may compute $M_j^{ab}{}_{:ab}$ in spherical polar coordinates by
substituting Eq.~(\ref{Msimple}) into Eq.~(\ref{MderivsG}), obtaining
\begin{equation}
     M_{(j)}^{ab}{}_{:ab} = \left(1-x^2\right)P''_j(x)
         -4xP'_j(x) -2P_j(x),\qquad\qquad x\equiv\hatk\cdot\hatn.
\end{equation}
This can be simplified using the definition of the associated Legendre
polynomials $P^m_l(x)$ and the recursion relation Eq.~(\ref{Plmrecursion}),
giving 
\begin{equation}
     M_{(j)}^{ab}{}_{:ab} = -P^2_j(x) -2(j^2+j+1)P_j(x).
\end{equation}
$P^2_j(x)$ can be represented as a finite series in $P_l(x)$ using the integral
Eq.~(\ref{plpl2int}), giving finally
\begin{equation}
M^{ab}_{(j)}{}_{:ab}=\,-\sum_{l=0}^j (l+1)(l+2)c_{lj} P_l(\hatk\cdot\hatn),
    \qquad\qquad c_{lj}\equiv
       \cases{                       1              &$l=j$                \cr
              \displaystyle{2(2l+1)\over(l+1)(l+2)} &$l-j$ even and $l<j$ \cr
\noalign{\vskip2pt}
                                     0              &$l-j$ odd or $l>j$   \cr}
\label{theanswer}
\end{equation}
Making use of Eq. (\ref{PYconvolution}) and orthonormality of the spherical
harmonics,
\begin{equation}
     \int d\hatn\, Y_{(lm)}^*(\hatn)\, M_{(j)}^{ab}{}_{:ab}(\hatn)
     =\,-(l+1)(l+2)c_{lj} \sqrt{{ 4\pi \over 2l+1}} \, \delta_{m0},
\label{Mmoments}
\end{equation}
and
\begin{equation}
     a_{(lm)}^{{\rm G,s}}(\veck)= {1\over 4} \sqrt{\pi(2l+1)}
     \, \delta_{m0}\, \Delta_{{\rm G}\,l}^{\rm s}(k),
\label{scalarGalm}
\end{equation}
where
\begin{equation}
     \Delta_{{\rm G}\,l}^{\rm s}(k) \equiv  
     -\left({2(l+1)(l+2)\over l(l-1)}\right)^{1/2}
     \sum_{j=l}^\infty
     \left( {2j+1 \over 2l+1} \right) c_{lj} \Delta_{Qj}^{\rm s}(k).
\label{infinitesum}
\end{equation}
This infinite sum is, as shown below, equivalent to the finite sum of
Eq.~(\ref{finitesum}). 
The contribution to $C_l^{\rm G}$ from this $\veck$
mode is
\begin{equation}
     C_l^{{\rm G,s}}(\veck) = {1\over 2l+1} \sum_{m=-l}^l
     |a_{(lm)}^{{\rm G,s}}(\veck)|^2 = {\pi \over 16}
     |\Delta_{{\rm G}\,l}^{\rm s}(k)|^2.
\end{equation}
The moments $C_l$ are rotationally invariant, so assuming statistical isotropy,
which guarantees that the different ${\bf k}$ modes are uncorrelated, the total
contribution of all scalar modes to $C_l^{\rm G}$ is
\begin{equation}
     C_l^{\rm G,scalar} = \int {d^3\veck \over (2\pi)^3}\,
     C_l^{{\rm G,s}}(\veck) = {1 \over 32 \pi} \int \, k^2 \,
     dk \, |\Delta_{{\rm G}\,l}^{\rm s}(k)|^2.
\label{ClGscalarresult}
\end{equation}

The calculation of $C_l^{\rm C,s}$ is similar with the replacements
$\cP^{ab}{}_{:ab} \rightarrow\cP^{ab}{}_{:ac}\epsilon^c{}_b$ and
$M_{(j)}^{ab}{}_{:ab} \rightarrow M_{(j)}^{ab}{}_{:ac}\epsilon^c{}_b$
in Eq. (\ref{almE}) [cf., Eqs. (\ref{Gmomentseasy}) and (\ref{Cmomentseasy})].
However substituting Eq.~(\ref{Msimple}) into  Eq.~(\ref{MderivsC}) yields
$M_{(j)}^{ab}{}_{:ac}\epsilon^c{}_b=0$ since $M_{(j)}^{ab}$ is diagonal
and independent of $\phi$.  This is just what we expect: Since
$M_{(j)}^{ab}$ is even under parity while $\epsilon^c{}_b$ is odd, the
product must integrate to zero.  Thus for scalar perturbations,
$a^{{\rm C,s}}(\veck)_{(lm)}=0$ and
\begin{equation}
     C_l^{\rm C,scalar}=0,
\end{equation}
as argued above.

\subsubsection{Polarization in the $\pm{\hat{\bf k}}$ directions}

\begin{figure}
\centerline{\psfig{figure=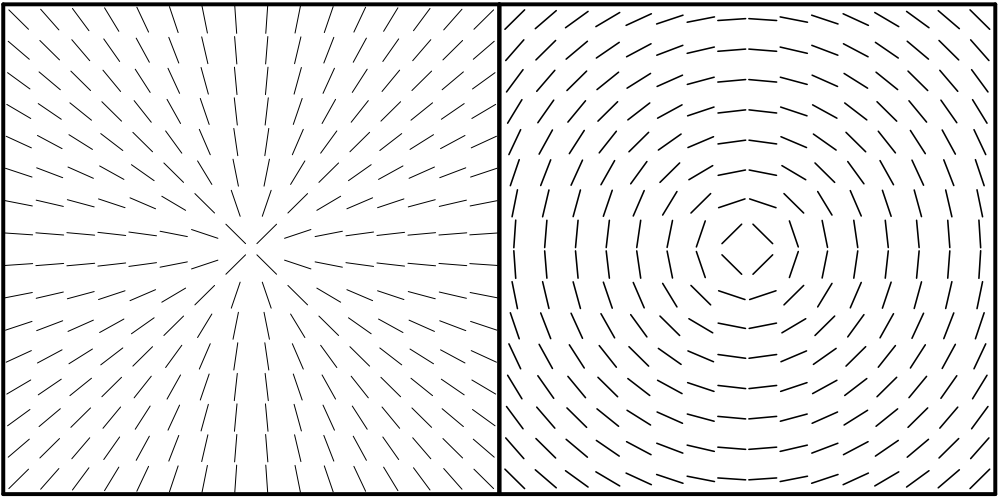,width=16.5cm}}
\caption{The basis tensors $M_{(j)}^{ab}(k)$ traditionally used for
polarization are discontinuous in the orientation of the polarization in the
directions on the sky both parallel and anti-parallel to  ${\bf k}$.  The
discontinuity is either as depicted in the left panel or as depicted in the
right panel. Note that we have just switched the sign of the polarization
between the two panels.}
\label{Discontinuous}
\end{figure}

As seen from Eq.~(\ref{theanswer}), when $l$ is large the coefficients
$c_{lj}$ at $l=j$ are 
much larger than ``nearby'' coefficients, say when $l=j-2$.
If the $l=j$ term dominates, then in the small angle approximation, i.e.,
for $l\gg1$, the approximation
$\Delta_{{\rm G}\,l}^{\rm s}(k)\approx\sqrt{2}\,\Delta_{Ql}^{\rm s}(k)$
is valid. However, this is {\it a priori} not 
a very good approximation since the contribution from
the terms with $l+2\leq j\leq 2l$ to the sum of 
Eq.~(\ref{infinitesum}) comes to
nearly as much as the contribution from the $j=l$ term
(although some cancellation may result from sign changes in
$\Delta^{\rm s}_{Ql}$).
The explanation is 
that it takes the sum of a large number $M_{(j)}^{ab}(\hatn)$ to
represent $Y_{(l0)}^{{\rm G}ab}(\hatn)$. This behavior is expected 
for the simple
reason that while the $Y_{(lm)}^{{\rm G}ab}$'s are smooth functions, the
$M_{(j)}^{ab}$'s are not: for $\hatn=\pm{\hat{\bf k}}$, i.e.,
when $\theta=0$ and $\pi$ (which are singular points of the spherical polar
coordinate system), $M_{(j)}^{ab}$ does not go to zero (since $P_j(1)\ne0$
and $P_j(-1)\ne0$).  Instead the amplitude of the polarization approaches a
constant but its direction varies discontinuously as illustrated in
Fig.~\ref{Discontinuous}.  To represent this discontinuous behavior as a
superposition of smooth functions requires a large number of terms. 
In fact, the
only reason why the sum of Eq.~(\ref{theanswer}) does not contain an infinite
number of terms is because it includes either $l=0$ if $j$ is even or $l=1$ if
$j$ is odd, neither of which are part of the basis of harmonic STF tensors.

	While the $M_{(j)}^{ab}$ basis functions are discontinuous, the
physical polarization pattern must be continuous. The polarization will
in fact be zero in the directions 
$\hatn=\pm{\hat{\bf k}}$ as can be seen directly from
the form of the Boltzmann equations \cite{annals}.  Thus the solution of the
Boltzmann equation will obey
\begin{equation}
{\cal P}_{ab}({\bf k},+{\hat{\bf k}})\propto
    \sum_{j=0}^\infty      (2j+1)\Delta_{Qj}^{\rm s}(k)=0 \qquad
{\cal P}_{ab}({\bf k},-{\hat{\bf k}})\propto
    \sum_{j=0}^\infty(-1)^j(2j+1)\Delta_{Qj}^{\rm s}(k)=0
\end{equation}
which in turn implies
\begin{equation}
\sum_{j\ge0}^{\rm even}(2j+1)\Delta_{Qj}^{\rm s}(k)=
\sum_{j\ge1}^{\rm odd }(2j+1)\Delta_{Qj}^{\rm s}(k)=0 .
\label{ContinuityCondition}
\end{equation}
Assuming a continuous polarization pattern and substituting
Eqs.~(\ref{Pscalarpattern}) and (\ref{theanswer}) into Eq.~(\ref{Mmoments}) 
gives
\begin{eqnarray}
\int d\hatn\, Y_{(00)}^*(\hatn)\,{\cal P}^{ab}{}_{:ab}({\bf k},\hatn)
    &\propto&\sum_{j=0}^\infty(2j+1)\,c_{0j}\Delta_{Qj}^{\rm s}(k)
            =\sum_{j\ge0}^{\rm even}(2j+1)\,\Delta_{Qj}^{\rm s}(k)=0,    \cr
\int d\hatn\, Y_{(10)}^*(\hatn)\,{\cal P}^{ab}{}_{:ab}({\bf k},\hatn)
    &\propto&\sum_{j=0}^\infty(2j+1)\,c_{1j}\Delta_{Qj}^{\rm s}(k)
            =\sum_{j\ge1}^{\rm odd }(2j+1)\,\Delta_{Qj}^{\rm s}(k)=0.
\end{eqnarray}
Thus the decomposition of ${\cal P}^{ab}{}_{:ab}$
actually has no $l=0$ or $l=1$ content.

Define the coefficients
\begin{equation}
b_{lj}\equiv
       \cases{\displaystyle{-{l^2-1\over(l+1)(l+2)}}&$l=j$                 \cr
\noalign{\vskip2pt}
              \displaystyle{2(2l+1)\over(l+1)(l+2)} &$l-j$ even and $0<j<l$\cr
\noalign{\vskip2pt}
                                     0              &$l-j$ odd  or  $j>l$  \cr}
\end{equation}
which have the property
\begin{equation}
{(l+1)(l+2)\over2(2l+1)}\,\left(
       \sum_{j=0}^l      b_{lj}\,(2j+1)\,\Delta_{Qj}^{\rm s}(k)
      +\sum_{j=l}^\infty c_{lj}\,(2j+1)\,\Delta_{Qj}^{\rm s}(k)\right)=
\cases{\displaystyle{\sum_{j\ge0}^{\rm even}(2j+1)\Delta_{Qj}^{\rm s}(k)}
                                                                & $l$ even,\cr
       \displaystyle{\sum_{j\ge1}^{\rm odd }(2j+1)\Delta_{Qj}^{\rm s}(k)}
                                                                & $l$ odd. \cr}
\end{equation}
Since these sums are zero for smooth (cosmological) polarization patterns, 
we may use the equality
\begin{equation}
  \sum_{j=l}^\infty c_{lj}\,(2j+1)\,\Delta_{Qj}^{\rm s}(k)
=-\sum_{j=0}^l      b_{lj}\,(2j+1)\,\Delta_{Qj}^{\rm s}(k)
\end{equation}
in Eq.~(\ref{infinitesum}) to obtain
\begin{equation}
     \Delta_{{\rm G}\,l}^{\rm s}(k) =
     \left({2(l+1)(l+2)\over l(l-1)}\right)^{1/2}
     \sum_{j=0}^l
     \left( {2j+1 \over 2l+1} \right) b_{lj} \Delta_{Qj}^{\rm s}(k).
\label{finitesum}
\end{equation}
We have transformed an infinite sum into a finite sum. While these finite sums
are still somewhat cumbersome, they are significantly less complicated than
previous expressions for moments of Stokes parameters obtained 
in the small-angle limit \cite{annals,flatsky}.

The fact that continuity demands zero polarization at
$\hatn=\pm{\hat{\bf k}}$ is reflected in that both 
$Y^{\rm G}_{(l0)ab}$  and $Y^{\rm C}_{(l0)ab}$ are zero at $\theta=0$ and
$\pi$, or equivalently that $G^\pm_{(l0)}(\pm1)=0$; continuity in fact
requires $G^\pm_{(lm)}(\pm1)=0$ for all $m$ except when $m=\pm2$.  Since the
$m=\pm2$ terms correspond to tensor perturbations 
when ${\hat{\bf k}}$ and ${\hat{\bf z}}$ are aligned, the polarization
{}from tensor perturbations does not give large sums, as evidenced
below. However, a similar treatment of vector perturbations ($m=\pm1$) 
leads to the same sort of infinite sums encountered for scalar perturbations.

\subsubsection{Cross moments}

{}From Eqs. (\ref{scalarTalm}) and (\ref{scalarGalm}), the
contribution to the TG cross moments from a single scalar mode
in the $\hatk=\hatz$ direction is
\begin{equation}
     C_l^{{\rm TG,s}}(\veck) = {1\over 2l+1} \sum_{m=-l}^l
     \left(a_{(lm)}^{{\rm T,s}}(\veck) \right)^* a_{(lm)}^{{\rm G,s}}(\veck),
\end{equation}
so integrating over all $\hatk$ gives
\begin{equation}
     C_l^{\rm TG,scalar} = \int {d^3\veck \over (2\pi)^3}\,
     C_l^{{\rm TG,s}}(\veck) = {1 \over 16 \pi} \int \, k^2 \,
     dk \, \Delta_{{\rm G}\,l}^{\rm s}(k) \Delta_{Il}^{\rm s}(k).
\label{ClTGscalar}
\end{equation}
The vanishing of $a^{{\rm C,s}}_{(lm)}(\veck)$ also demonstrates
explicitly that the moments $C_l^{\rm TC,s}=C_l^{\rm GC,s}=0$,
as argued above.

\subsection{Tensor Metric Perturbations}

\subsubsection{Temperature moments}

The calculation of tensor moments proceeds in an analogous
fashion.  Recall, however, that tensor perturbations have
two polarization states, $+$ and $\times$.
Consider a single Fourier mode with $+$ polarization and as before
choose a coordinate system with $\bf\hat z$ in the $\hatk$ direction.
{}From Eq. (7.1) in Ref. \cite{annals}, the contribution of this
$\bf k$-mode to the temperature anisotropy is
\begin{equation} 
     T({\bf k},\hatn)_+ = {T_0\over 4}
     \sum_j (2j+1) P_j(\hatk\cdot\hatn) \,\sin^2\theta\,\cos 2\phi\,
     \tilde\Delta_{Il}^+(k),
\end{equation}
(note that the choice of the zero of
$\phi$ is arbitrary and inconsequential), and $\tilde\Delta_{Il}^+(k)$
is the perturbation to the photon brightness induced by this tensor
mode after the Polnarev change of variables \cite{polnarev}. For the $\times$
polarization state, simply replace $\cos2\phi$ with
$\sin2\phi$ and $\tilde\Delta_{Il}^+$ with $\tilde\Delta_{Il}^\times$.
Again, we expand this anisotropy pattern in spherical harmonics,
\begin{equation}
     a_{(lm)}^{{\rm T},+}(\veck) = {1\over4} \sum_j (2j+1)
     \tilde\Delta_{Ij}^+(k)
     \int d\hatn \, P_j(\hatk\cdot\hatn) Y_{(lm)}^*(\hatn)
     \sin^2\theta\, \cos2\phi.
\label{tensoralmT}
\end{equation}
Note that for
$\hatk={\bf \hat z}$, $P_j(\cos\theta) = [4\pi/(2l+1)]^{1/2}
Y_{(j0)}$.  Then Eq.~(\ref{iteratedone}) can be used
to express the integrand in Eq.~(\ref{tensoralmT}) as a sum of
products of two spherical harmonics.  Orthonormality of
spherical harmonics then gives 
\begin{eqnarray}
     a_{(lm)}^{{\rm T},+}(\veck) = {1\over 8} &&
     (\delta_{m,2}+\delta_{m,-2}) \sum_j (2j+1) \sqrt{{ 4\pi
     \over 2j+1}} \Delta_{Ij}^+(k) \Biggl\{ 
     \left[ {(j+1)(j+2)(j+3)(j+4) \over (2j+1)(2j+3)^2(2j+5) }
     \right]^{1/2} \delta_{l,j+2} \nonumber \\
     && + {2 \sqrt{j(j+1)(j-1)(j+2)} \over (2j-1)(2j+3)} \delta_{lj}
     +\left[ {j(j-1)(j-2)(j-3) \over (2j-1)^2(2j+1)(2j-3) }
     \right]^{1/2} \delta_{l,j-2} \Biggr\}.
\end{eqnarray}
The $\delta_{lx}$'s project out only three terms in the sum, and
this reduces to
\begin{equation}
     a_{(lm)}^{{\rm T},+}(\veck) = {1\over 4}\sqrt{\pi(2l+1)}
     (\delta_{m2}+\delta_{m,-2}) \sqrt{{(l+2)! \over (l-2)!}}
     \left[ {\Delta_{I,l-2}^+(k) \over (2l-1)(2l+1)} - {2
     \Delta_{Il}^+(k) \over (2l+3)(2l-1)} + {\Delta_{I,l+2}(k)
     \over (2l+3)(2l+1)} \right].
\label{almTtensors}
\end{equation}
For the $\times$ polarization state, simply replace
$(\delta_{m2}+\delta_{m,-2})$ with 
$-i(\delta_{m2}-\delta_{m,-2})$, and $\Delta^+$ with
$\Delta^\times$.

The contribution of this $\veck$ mode to $C_l^{\rm T}$ is
then 
\begin{eqnarray}
     C_l^{{\rm T},+}(\veck) &=& {1\over 2l+1} \sum_m
     |a_{(lm)}^{{\rm T},+}(\veck)|^2 \nonumber \\
     &=& {\pi \over 8} { (l+2)! \over (l-2)!}
     \left[ {\Delta_{I,l-2}^+(k) \over (2l-1)(2l+1)} - {2
     \Delta_{Il}^+(k) \over (2l+3)(2l-1)} + {\Delta_{I,l+2}(k)
     \over (2l+3)(2l+1)} \right]^2,
\end{eqnarray}
and the result for the $\times$ polarization state is
the same (with the replacement $+\rightarrow\times$, of course).
If the spectrum of $+$ and $\times$ states is the same (which is
demanded by statistical isotropy), then
the total contribution of tensor modes to the temperature
anisotropy is
\begin{eqnarray}
     C_l^{{\rm T},{\rm tensor}} & = & 2\int {d^3\veck \over (2\pi)^3} \,
     C_l^{{\rm T},+}(\veck) \nonumber \\
     & = & {1\over 8\pi} {(l+2)! \over (l-2)!} \,\int k^2 
     dk \, 
     \left[ {\Delta_{I,l-2}^+(k) \over (2l-1)(2l+1)} - {2
     \Delta_{Il}^+(k) \over (2l+3)(2l-1)} + {\Delta_{I,l+2}(k)
     \over (2l+3)(2l+1)} \right]^2,
\label{ClTtensors}
\end{eqnarray}
which agrees with the results of previous calculations
\cite{annals,tensors,aw84}. 

\subsubsection{Polarization Moments}

Finally, we calculate the polarization moments from tensor
perturbations. The Stokes parameters induced by
a single tensor Fourier mode with $+$ polarization 
in the direction $\hatn=(\theta,\phi)$ are \cite{annals}
\begin{eqnarray}
     Q({\bf k},\hatn)_+ &=& {T_0\over4} \sum_j (2j+1)P_j(\cos\theta)
     (1+\cos^2\theta)\cos 2\phi\,\tilde\Delta_{Qj}^+(k),\\
     U({\bf k},\hatn)_+ &=& {T_0\over4} \sum_j (2j+1)P_j(\cos\theta)
     2\cos\theta\sin 2\phi\,\tilde\Delta_{Qj}^+(k),
\end{eqnarray}
where again $\tilde\Delta_{Ql}^{\rm s}(k)$ are Legendre coefficients of
the photon polarization brightness for tensor metric perturbations
with $+$ polarization. Note $U$ is the opposite sign from Ref.~\cite{annals}
because the coordinate system there has opposite orientation from
the one here. The polarization tensor is thus
\begin{equation}
     \cP^{ab}_+(\veck,\hatn) ={T_0\over8} \sum_j (2j+1)\,
     \Delta_{Qj}^+(k)\, M_j^{ab}(\hatn),
\label{sevenfour}
\end{equation}
with
\begin{equation}
     M_{(j)}^{ab}(\hatn) = P_j(\cos\theta)\left( 
     \begin{array}{cc}
     (1+\cos^2\theta)\cos2\phi & -2 \cot\theta \sin 2\phi\\
     \noalign{\vskip6pt}
     -2 \cot\theta \sin 2\phi &
     -(1+\cos^2\theta)\csc^2\theta\cos2\phi\\
     \end{array} \right).
\label{xipluseqn}
\end{equation}
For $\times$ polarization, make
the replacements $\cos2\phi \rightarrow \sin
2\phi$ and $\sin2\phi \rightarrow -\cos 2\phi$.

Now we calculate the multipole moments induced by this single
tensor mode.  From Eq. (\ref{Gmomentseasy}) above,
\begin{equation}
     a_{(lm)}^{{\rm G},+}(\veck) 
     = {N_l \over 8} \sum_j (2j+1) \tilde\Delta_{Qj}^+(k) \,
     \int d\hatn\, Y_{(lm)}^*(\hatn) \,
     M_{(j)}^{ab}{}_{:ab}(\hatn).
\label{almEplusveck}
\end{equation}
Calculation of this $M_{(j)}^{ab}{}_{:ab}$ is straightforward using
Eq.~(\ref{MderivsG}), but the algebra is considerably longer
than the scalar case.  The result is
\begin{equation}
     M_{(j)}^{ab}{}_{:ab} = \cos2\phi \left[ 12(1-x^2) P_j(x) 
     + 8x(1-x^2)P_j'(x) + (1-x^4) P_j''(x)\right],
\label{Mjababeqn}
\end{equation}
with $x=\cos\theta$.
The $\times$ perturbations
give the same result with $\cos2\phi\rightarrow \sin
2\phi$.  Because of this azimuthal dependence, the integral
in Eq.~(\ref{almEplusveck}) is non-zero only for
$m=\pm 2$.  Using Eqs. (\ref{pmldef}), (\ref{plmidentity2}), and
(\ref{parttwo}),  we get
\begin{equation}
     M_{(j)}^{ab}{}_{:ab} = \cos2\phi \left[ {(j+3)(j+4) P_{j+2}^2 \over
     (2j+1)(2j+3) } + {6j(j+1)P_j^2 \over (2j+3)(2j-1)} + {(j-2)(j-3)
     P_{j-2}^2 \over (2j+1)(2j-1)} \right].
\end{equation}
This is just a sum of the three spherical harmonics with
$m=2$ and the three with $m=-2$.  The integral in
Eq. (\ref{almEplusveck}) is
\begin{equation}
     \int d\hatn Y_{(lm)}(\hatn) M_{(j)}^{ab}{}_{:ab}(\hatn) = \sqrt{ {\pi
     \over 2l+1}{ (l+2)! \over (l-2)!} } (\delta_{m2}+\delta_{m,-2})
     \left[ {(l+1)(l+2)
     \delta_{l,j+2} \over (2l-3)(2l-1)} + {6l(l+1) \delta_{lj} \over
     (2l+3)(2l-1)} + {l(l-1) \delta_{l,j-2} \over (2l+5)(2l+3) }
     \right],
\end{equation}
which results in
\begin{equation}
     a_{(lm)}^{{\rm G},+}(\veck) = {1\over8} (\delta_{m2}+\delta_{m,-2})
     \sqrt{2\pi (2l+1)} \left[ {(l+2)(l+1) \Delta_{Q,l-2}^+(k) \over
     (2l-1)(2l+1)} + {6l(l+1) \Delta_{Ql}^+(k) \over (2l+3)(2l-1)} +
     {l(l-1) \Delta_{Q,l+2}^+(k) \over (2l+1)(2l+3)} \right].
\label{almGtensors}
\end{equation}
As before, each $\veck$ mode and each
polarization state contributes to $C_l$ in the same way, so
integrating over $\veck$ and multiplying by two (for the two
polarization states) gives
\begin{equation}
     C_l^{{\rm G},{\rm tensor}} = {1\over 16\pi} \int k^2 dk \left[
     {(l+2)(l+1) \Delta_{Q,l-2}^+(k) \over
     (2l-1)(2l+1)} + {6l(l+1) \Delta_{Ql}^+(k) \over (2l+3)(2l-1)} +
     {l(l-1) \Delta_{Q,l+2}^+(k) \over (2l+1)(2l+3)} \right]^2. 
\label{correctClE}
\end{equation}
Therefore, $C_l^{{\rm G},{\rm tensor}}$ is due to three
Legendre expansion coefficients, $\Delta_{Ql}^+(k)$, as opposed
to an infinite sum as in the scalar case.

The derivation of the multipole moments for the C harmonics
is similar. From Eq. (\ref{Cmomentseasy}) above,
\begin{equation}
     a_{(lm)}^{{\rm C}\,+}(\veck) 
     = {N_l \over 8} \sum_j (2j+1) \tilde\Delta_{Qj}^+(k) \,
     \int d\hatn\, Y_{(lm)}^*(\hatn) \,
     M_{(j)}^{ab}{}_{:ac}(\hatn)\epsilon^c{}_b.
\label{almBplusveck}
\end{equation}
This time we get for the ``+'' gravity wave polarization
\begin{eqnarray}
     M_{(j)}^{ab}{}_{:ac}\epsilon^c{}_b & = & \sin 2\phi \left[  
      -8(1-x^2)P_j'(x) - 2x(1-x^2) P_j''(x)\right]\nonumber\\
      & = & \sin 2\phi \left[-{2(j+3)\over 2j+1}P^2_{j+1}(x)
                             -{2(j-2)\over 2j+1}P^2_{j-1}(x)\right]
\label{Mjababeqntimes}
\end{eqnarray}
where the second equality uses the identities (\ref{plmidentity2})
and (\ref{plmidentity1}). 
For $\times$ polarization states,
replace $\sin2\phi\rightarrow -\cos 2\phi$.  
The multipole coefficients are then
\begin{equation}
     a^{{\rm C},+}(\veck)_{lm} = -{i\over 4}(\delta_{m2}-\delta_{m-2})
\sqrt{2\pi\over 2l+1} \left[ (l+2)\tilde\Delta^+_{Ql-1}(k) +
(l-1)\tilde\Delta^+_{Ql+1}(k)\right].
\end{equation}
The $\times$ perturbations give the same result except for a minus
sign between the Kronecker delta's and an overall factor of $i$.
Again assuming equal contributions from both $+$
and $\times$ tensor perturbations,
\begin{equation}
      C_l^{{\rm C},{\rm tensor}}={1\over 4\pi}\int dk\, k^2
      \left[{l+2\over 2l+1} \tilde\Delta^+_{Ql-1}(k)
      + {l-1\over 2l+1}\tilde\Delta^+_{Ql-1}(k)\right]^2.
\label{ClBplusk}
\end{equation}
This calculation verifies our qualitative arguments that tensor
modes will produce a ``C'' polarization field.

\subsubsection{Cross Moments}

{}From Eqs. (\ref{almTtensors}) and (\ref{almGtensors}), 
the non-zero cross moments are
\begin{eqnarray}
     C_l^{{\rm TG},{\rm tensor}} & = & {1\over 8\pi N_l}\int k^2 dk
     \left[ {\tilde\Delta_{I,l-2}^+(k) \over
     (2l-1)(2l+1)} - {2 \tilde\Delta_{Il}^+(k) \over (2l+3)(2l-1)} +
     {\tilde\Delta_{I,l+2}^+(k) \over (2l+3)(2l+1)} \right] \nonumber \\
     & \times & \left[ {(l+2)(l+1) \tilde\Delta_{Q,l-2}^+(k) \over
     (2l-1)(2l+1)} + {6l(l+1) \tilde\Delta_{Ql}^+(k) \over (2l+3)(2l-1)} +
     {l(l-1) \tilde\Delta_{Q,l+2}^+(k) \over (2l+1)(2l+3)} \right],
\label{ClTGtensors}
\end{eqnarray}
assuming equal contributions from the $+$ and $\times$ tensor
perturbations. Note the $N_l^{-1}$ in the prefactor comes from the
temperature coefficient, not from the polarization coefficient. 

Both the temperature and G multipole coefficients for a $\veck$
tensor mode with $+$ polarization,
$a_{(lm)}^{{\rm T},+}(\veck)$ and $a_{(lm)}^{{\rm G},+}(\veck)$
are proportional to $(\delta_{m2}+\delta_{m,-2})$ (and similarly
for $\times$ modes with the sum replaced by a difference.  On
the other hand, the corresponding C multipole coefficients, 
$a_{(lm)}^{{\rm C},+}(\veck)$ are proportional to
$(\delta_{m2}-\delta_{m,-2})$ (and the same with the difference
replaced by a sum for $\times$ modes).  Therefore, after summing
over $m$, the cross moments $C_l^{{\rm TC},{\rm tensor}}$ and
$C_l^{{\rm GC},{\rm tensor}}$ for tensor perturbations vanish.
This is a consequence of the symmetry arguments mentioned above.

\subsection{Summary}

In this Section we have calculated the CMB temperature and
polarization multipole moments for both scalar and tensor
perturbations.  For theories in which the perturbations have a
Gaussian distribution (such as inflationary models), this set of
multipole moments fully specifies the statistical properties of
the combined temperature/polarization map.  In virtually all
theories considered, scalar and tensor perturbations are
statistically independent, so their contributions to the
CMB temperature anisotropy and polarization add in
quadrature.  Even if they are not statistically independent,
angular orthogonality of the different modes
(i.e., scalar, vector, and tensor modes in the $\hatz$ direction
induce non-zero multipole coefficients only for $m=0$, $m=\pm$,
and $m=\pm2$, respectively) guarantees that the contributions of
scalar, vector, and tensor perturbations to the multipole
moments will add in quadrature.  Therefore, for Gaussian
theories, the map will be fully
described by $C_l^{\rm T}=C_l^{\rm T,scalar}+C_l^{{\rm T},{\rm tensor}}$
[cf., Eqs. (\ref{ClTs}) and (\ref{ClTtensors})], 
$C_l^{\rm G}=C_l^{\rm G,scalar}+C_l^{{\rm G,tensor}}$ 
[cf., Eqs. (\ref{ClGscalarresult}) and (\ref{correctClE})],
$C_l^{\rm C}=C_l^{{\rm C,tensor}}$
[cf., Eqs. (\ref{ClBplusk})], and
$C_l^{\rm TG}=C_l^{\rm TG,scalar}+C_l^{{\rm TG,tensor}}$ 
[cf., Eqs. (\ref{ClTGscalar}) and (\ref{ClTGtensors})].  For
non-Gaussian theories, $n$-point correlation functions with $n>2$
may be nonzero.

Eq. (\ref{ClTs}) for the temperature moments from scalar
perturbations is written as an integral of a square of a single
$\Delta_{Il}$.  However, all of the other moments are written as
squares of a sum of several $\Delta_{Il}$'s and/or
$\Delta_{Ql}$'s.  This is because a
spherical-harmonic decomposition is not natural for a Stokes
parameter $Q$, which is a tensor component, nor for the
brightness perturbation from tensor modes.  The calculation can
be reformulated using a tensor-harmonic expansion for these
quantities, which provide a natural basis
\cite{nextpaper}.

\subsection{Line-of-Sight Approach}

A very efficient and accurate algorithm for computation of
multipole moments has recently been given by Seljak and
Zaldarriaga \cite{sz96a} and applied to the polarization
multipole moments \cite{sz96c} (hereafter, SZ).  Although 
significantly different in appearance, 
their spin-harmonic formalism is equivalent to
our tensor-harmonic formalism.  Here, we briefly compare our
results with theirs.  Although the formalisms differ and the
calculations are somewhat lengthy, we find that the results agree,
which gives us confidence in both sets of results.
Furthermore, by identifying the moments in their paper with
those in ours, their numerical code (which has been
made publicly available) can be used
to compute the multipole moments presented in this paper.

Consider, for example, the G polarization moments from scalar
perturbations.  According to Eqs. (12) and (14) in SZ, the
polarization pattern induced on the sky by a scalar $\veck$ mode
in the $\hatz$ direction (i.e., their version of our
Eq.~(\ref{QUscalar})) is
\begin{equation}
     Q(\hatn) = {3\over4}(1-\cos^2\theta) \int_0^{\tau_0}\,
     d\tau\, e^{ik\tau\cos\theta} g(\tau) \Pi(k,\tau) \xi(\veck),
\end{equation}
and $U(\hatn)=0$, where $\tau$ is the conformal time, $g(\tau)$
is a visibility function, $\Pi$ is a combination of
intensity and polarization perturbations, and $\xi(\veck)$ is an
amplitude for the scalar mode (see SZ for more details).
Therefore, the polarization tensor induced by this scalar mode
is
\begin{equation}
     P^{ab}(\hatn) = {3\over4} \xi(\veck) \int_0^{\tau_0}\,
     d\tau\, g(\tau) \Pi(k,\tau) e^{ik\tau\cos\theta} \left( 
     \begin{array}{cc}
     1 & 0 \\
     0 & -\csc^2\theta \\
     \end{array} \right).
\end{equation}
Using the rules of covariant differentiation, we find that 
\begin{eqnarray}
     P^{ab}{}_{:ab}(\hatn) & =& {3\over8} \xi(\veck) \int_0^{\tau_0}\,
     d\tau\, g(\tau) \Pi(k,\tau)  \left( {\partial^2 \over
     \partial\theta^2} + 3 {\cos\theta \over \sin\theta}
     {\partial \over \partial\theta} -2 \right)
     e^{ik\tau\cos\theta} \nonumber \\
     &=& {3\over8} \xi(\veck) \int_0^{\tau_0}\,
     d\tau\, g(\tau) \Pi(k,\tau)  
     { d^2 \over d(\cos\theta)^2}
     \left[ (1-\cos^2\theta)^2 e^{ik\tau\cos\theta}\right],
\end{eqnarray}
which agrees with their $\Delta_{\tilde E}^{(S)}/2$ given in their
Eq. (15).  The $C_l^{\rm G,scalar}$ moments are obtained by
plugging this into Eq. (\ref{Gmomentseasy}), squaring,
summing over $m$ and then integrating over $\veck$.  We then
find that our results agree with theirs (realizing that our
$(2\pi)^3$ Fourier conventions differ from theirs) if we
identify $C_l^{\rm G}= C_{El}/2$, where $C_{El}$ are their
electric-type moments.  We have further checked that our
temperature moments agree with theirs (with no factor of two),
and that our C moments are twice their $B$ moments: $C_l^{\rm C}
= C_{Bl}/2$.  Our $C_l^{\rm TG}$ are equal to their
$C_{Cl}/\sqrt{2}$.  Although we do not present it here, we have
checked their tensor-mode calculations as well.  Although significantly
more involved, we still find agreement; for example, compare
their ${\cal E}(x)$ and ${\cal B}(x)$ with our
Eqs. (\ref{Mjababeqn}) and (\ref{Mjababeqntimes}).  The
identification of our polarization moments with theirs 
is also consistent with our $w_P^{-1}$ in Eq.~(\ref{winverse}) 
being half theirs.

\section{Two-Point Correlation Functions}

In this Section, we relate the multipole moments
$C_l^{\rm T}$, $C_l^{\rm G}$, $C_l^{\rm C}$, and $C_l^{\rm TG}$ 
to two-point correlation functions of temperature and Stokes 
parameters.  This will make contact with previous work on the subject.
We also derive flat-sky limits useful for analyzing maps of small
sky patches.

\subsection{Correlations Between Temperature and Stokes Parameters}

The linear-polarization state at any given point is specified
completely by the Stokes parameter $Q$ and $U$, but these quantities
depend on the coordinate system which one chooses.  On the other hand,
we know that $Q$ and $U$ transform as the components of a STF
$2\times2$ tensor, so given $Q$ and $U$ in some coordinate system, we
can always determine $Q'$ and $U'$ in any other coordinate system.

The Universe is assumed to be statistically isotropic, so it is possible
to construct two-point correlation functions which depend only on
the angular separation between the two points.  But simply correlating
$Q$ and $U$ in a particular coordinate system
gives correlation functions which depend on the positions of the
points being correlated as well the angular separation.  This is
what has been done in previous published work.

A coordinate-independent set of correlation functions 
can be expressed in terms of the ones which have appeared in
the previous literature.  The prescription is simply to define
correlation functions of Stokes parameters with respect to
axes which are parallel and perpendicular to the great arc (or
geodesic) connecting the two points being correlated.  So $Q_r$ is
the difference in intensities in two linear-polarization states parallel
and perpendicular to the great arc connecting the two points, and $U_r$
is the difference in two linear-polarization states which lie
$45^\circ$ away from the parallel and perpendicular.
The three quantities $T$, $Q_r$, and $U_r$
have six correlation functions between them: 
$\VEV{TT}$, $\VEV{U_rU_r}$, $\VEV{Q_rQ_r}$,
$\VEV{Q_rT}$, $\VEV{Q_rU_r}$, and $\VEV{U_rT}$.  However,
only four can be nonzero.  Although $Q_r$ and $T$ are invariant
under reflection along the great arc connecting the two points being
correlated, $U_r$ changes sign. 
Therefore, the expectation values $\VEV{Q_rU_r}$
and $\VEV{U_rT}$ must be zero from statistical isotropy.  This is
as expected:  four nonzero sets of
moments $C_l^{\rm T}$, $C_l^{\rm G}$, $C_l^{\rm C}$, and
$C_l^{\rm TG}$ describe the map.  Correspondingly, 
four nonzero correlation functions provide
an equivalent statistical description.

We begin with the familiar $\VEV{TT}$ correlation function,
\begin{equation}
     C^{\rm T}(\theta) = \VEV{ {T(\hatnone) \over T_0} 
     {T(\hatntwo) \over T_0}}_{\hatnone\cdot \hatntwo=\cos\theta}.
\label{Tcorrelation}
\end{equation}
The correlation function depends only on the angular separation of the
two points, so in calculating it, we may choose one point to be at the
north pole, $(0,0)$, and the other to be on the $\phi=0$ longitude at
a distance $\theta$ from the north pole, $(\theta,0)$.  Then expand
$T(\hatn)$ in terms of spherical harmonics as in
Eq.~(\ref{Texpansion}) and note that $Y_{(lm)}(0,0) =
\delta_{m0}\sqrt{(2l+1)/(4\pi)}$. So
\begin{eqnarray}
     C^{\rm T}(\theta) &=& \VEV{ {T(0,0) \over T_0}
     {T(\theta,0) \over T_0}} \nonumber \\
     & =& \sum_{lml'm'} \VEV{a_{(lm)}^{{\rm T}\,*} a_{l'm'}^{\rm T}} Y_{(lm)}^*(0,0)
       Y_{(l'm')}(\theta,0) \nonumber \\
     & =& \sum_{lml'm'} C_l^{\rm T}  \delta_{ll'} \delta_{mm'}  \sqrt{ {2l+1
     \over 4 \pi} } \delta_{m0} Y_{(l'm')}(\theta,0) \nonumber \\
     & =& \sum_l\, {2l+1 \over 4 \pi}\, C_l^{\rm T}\, P_l(\cos\theta),
\label{TTresult}
\end{eqnarray}
where we have used Eq.~(\ref{cldefs}) to go from the second
to the third line.  This recovers the well-known result for the
temperature autocorrelation function.

The derivation of the polarization correlation functions will proceed
analogously (and is similar to the case for weak-lensing
correlation functions \cite{albert}), and requires $W_{(lm)}$
and $X_{(lm)}$ at the north pole.  
Using the asymptotic relations
\begin{equation}
     P_l^m(\cos\theta) \sim { (-1)^{(m+|m|)/2} \over 2^{|m|} |m|!}
     {(l+|m|)! \over (l-|m|)!} \, \theta^{|m|}, \qquad 
     \theta\rightarrow0 \qquad\qquad (m\neq 0)
\end{equation}
\begin{equation}
P_l(\cos\theta)\sim1-{1\over 4}l(l+1)\theta^2, \qquad \theta\rightarrow 0
\end{equation}
it is straightforward to show that for $|m|\geq 2$,
$X_{(lm)}$ and $Y_{(lm)}$ are
both asymptotic to $\theta^{|m|-2}$ as $\theta\rightarrow0$,
so they are nonzero
at $\theta=0$ only for $|m|=2$; for $m=0$ and $m=1$, 
they are asymptotic to $\theta^2$ and $\theta$. 
After a little algebra, 
\begin{equation}
     W_{(lm)}(0,0) = {1\over2} \sqrt{ {2l+1 \over 4\pi} { (l+2)! \over
     (l-2)!} } ( \delta_{m2} + \delta_{m,-2}),
\label{Wzero}
\end{equation}
and
\begin{equation}
     X_{(lm)}(0,0) = {i\over2} \sqrt{ {2l+1 \over 4\pi} { (l+2)! \over
     (l-2)!} } ( \delta_{m2} - \delta_{m,-2}).
\label{Xzero}
\end{equation}

Now consider the $\VEV{QQ}$ correlation function
\begin{equation}
     C^Q(\theta) = \VEV{ {Q_r(\hatnone)\over T_0}
      {Q_r(\hatntwo)\over T_0} }_{\hatnone\cdot
     \hatntwo=\cos\theta},
\label{Qcorrelation}
\end{equation}
where, once again, the Stokes parameters $Q_r$ are defined with
respect to axes parallel and perpendicular to the great arc
connecting $\hatnone$ and $\hatntwo$.  As in the temperature case, 
choose one point to be at the north pole and another a distance
$\theta$ away along the $\phi=0$ longitude.  This choice has the added
advantage that the great arc connecting these two points is along the
$\theta$ direction, so we can use the $Q$ defined in the
$(\hat\theta,\hat\phi)$ coordinate system,
Eq. (\ref{QUexpansion}).  However, in this coordinate
system, the definition of $Q$ at the north pole is, strictly speaking,
ambiguous.  Therefore, we always
consider a
point on the $\phi=0$ longitude which is infinitesimally close to the
north pole;  in other words, $Q(0,0)$ really means
${\rm lim}_{\theta\rightarrow0} Q(\theta,0)$. 
Using Eq.~(\ref{QUexpansion}) for the Stokes parameters, the $\VEV{QQ}$
correlation function is then
\begin{eqnarray}
     C^Q(\theta) &=& \VEV{ {Q(0,0)\over T_0} 
       {Q(\theta,0)\over T_0} } \nonumber \\
      &=& \sum_{lml'm'} N_l N_{l'}
     \VEV{  [a_{(lm)}^{\rm G} W_{(lm)}(0,0) -
     a_{(lm)}^{\rm C} X_{(lm)}(0,0)] [a_{(l'm')}^{{\rm G}\,*}
     W_{(l'm')}^*(\theta,0) -
     a_{(l'm')}^{{\rm C}\,*} X_{(l'm')}^*(\theta,0)] } \nonumber \\
     & = & \sum_l \sqrt{2l+1 \over 8\pi}N_l [ C_l^{\rm G}(W_{(l2)}^* +
      W_{(l,-2)}^*) + iC_l^{\rm C}(X_{(l2)}^* - X_{(l,-2)}^*) ],
\end{eqnarray}
where we have used Eqs.~(\ref{cldefs}), (\ref{Xzero}), and (\ref{Wzero}), and
$\VEV{a_{(lm)}^{{\rm G}\,*}a_{(lm)}^{\rm C}}=0$.  This can be simplified using
$X_{(lm)}^* = -X_{(l,-m)}^*$ and $W_{(lm)}^* = W_{(l,-m)}^*$
and the definitions in Eqs.~(\ref{WGplus}) and (\ref{XGminus}), giving
\begin{equation}
     C^Q(\theta) = -\sum_l {2l+1 \over 2\pi} N_l^2 
     [ C_l^{\rm G} G^+_{(l2)}(\cos\theta)
     + C_l^{\rm C} G^-_{(l2)}(\cos\theta)].
\label{QQresult}
\end{equation}

For the $\VEV{UU}$ correlation function, the derivation is similar,
giving
\begin{equation}
     C^U(\theta) = -\sum_l {2l+1 \over 2\pi} N_l^2 
     [ C_l^{\rm C} G^+_{(l2)}(\cos\theta)
     + C_l^{\rm G} G^-_{(l2)}(\cos\theta)].
\label{UUresult}
\end{equation}
For the $\VEV{TQ}$ cross-correlation function,
\begin{eqnarray}
     C^{\rm TQ}(\theta) & = & \VEV{{T(\hatnone)\over T_0} 
                               {Q_r(\hatntwo)\over T_0} }_{\hatnone \cdot
     \hatntwo=\cos\theta} \nonumber \\
	& =& \sum_l\, {2l+1 \over 4\pi}\, N_l \, C_l^{\rm TG}\,
     P_l^2(\cos\theta).
\label{TQresult}
\end{eqnarray}

Eqs. (\ref{QQresult}), (\ref{UUresult}), and (\ref{TQresult}) are {\it
exact} (i.e., there is {\it no} small-angle approximation) expressions
for the polarization correlation functions.  

\subsection{Multipole Moments from Correlation Functions}

Above, we derived expressions for
correlation functions in terms of multipole moments, and now we
perform the inverse transform and express the multipole moments
in terms of correlation functions.
Begin with the temperature autocorrelation function:
multiply both sides of Eq. (\ref{TTresult}) by the Legendre
polynomial $P_{l'}(\cos\theta)$, integrate over $\cos\theta$, and
use the orthogonality of Legendre polynomials to obtain 
\begin{equation}
     C_l^{\rm T} = 2\pi \int_0^\pi d\theta\sin\theta
     P_l(\cos\theta)\, C^{\rm T}(\theta).
\label{TfromTT}
\end{equation}
Similarly, for the polarization-temperature moments,
multiply both sides of Eq. (\ref{TQresult}) by the associated
Legendre function $P_{l'}^2$, integrate, and use Eq.~(\ref{plmnorm})
to obtain
\begin{equation}
     C_l^{\rm TG} = \pi N_l \int_0^\pi d\theta\sin\theta
     P_l^2(\cos\theta)\, C^{\rm TQ}(\theta).
\label{TGfromTQ}
\end{equation}

The derivation of the polarization moments from the polarization
autocorrelation functions is similar. 
Orthonormality of the tensor harmonics implies that
\begin{equation}
     \int [ W_{(lm)}^*(\hatn) W_{(l'm')}(\hatn)
     + X_{(lm)}^*(\hatn) X_{(l'm')}(\hatn)]\, d\hatn = {2\over
     N_l^2} \delta_{ll'} \delta_{mm'},
\label{normalofXW}
\end{equation}
and it can also be shown that
\begin{equation}
     \int [ -X_{(lm)}^*(\hatn) W_{(l'm')}(\hatn)
     + W_{(lm)}^*(\hatn) X_{(l'm')}(\hatn)]\, d\hatn = 0.
\label{XWzero}
\end{equation}
To do so, note that the $\phi$ dependence of $W_{(lm)}$ and
$X_{(lm)}$ is just $e^{im\phi}$, which means that the integral
is immediately zero for $m\neq m'$.  For $m=m'$, 
the integral over $\cos\theta$ vanishes
using the explicit forms of $G_{(lm)}^\pm$ \cite{albert}. 
{}From Eqs. (\ref{QQresult}) and (\ref{UUresult}),
\begin{equation}
     [C^Q(\theta) +C^U(\theta)]\,e^{2i\phi} = \sum_l 
     \sqrt{2l+1 \over 2\pi} N_l
     (C_l^{\rm G} +C_l^{\rm C}) [W_{(l2)}(\theta,\phi) + i
     X_{(l2)}(\theta,\phi)] 
\label{QplusU}
\end{equation}
\begin{equation}
     [C^Q(\theta) -C^U(\theta)]\,e^{2i\phi} = \sum_l 
     \sqrt{2l+1 \over 2\pi} N_l
     (C_l^{\rm G} - C_l^{\rm C}) [W_{(l2)}(\theta,\phi) - i
     X_{(l2)}(\theta,\phi)].
\label{QminusU}
\end{equation}
Then multiply both sides of the first equation by $W_{(l2)}^* - i
X_{(l2)}^*$ and the second by $W_{(l2)}^* + i X_{(l2)}^*$,
integrate over all directions $\hatn$, and apply
Eqs. (\ref{normalofXW}) and (\ref{XWzero}), giving
\begin{equation}
     C_l^{\rm G} + C_l^{\rm C} = \sqrt{2\pi \over 2l+1} {N_l\over 2} \int
     d\hatn [C^Q(\theta)+C^U(\theta)]\, e^{2i\phi}\,
     [W_{(l2)}^*(\hatn) - i X_{(l2)}^*(\hatn)],
\label{partialone}
\end{equation}
\begin{equation}
     C_l^{\rm G} - C_l^{\rm C} = \sqrt{2\pi \over 2l+1} {N_l\over 2} \int
     d\hatn [C^Q(\theta)-C^U(\theta)]\, e^{2i\phi}\,
     [W_{(l2)}^*(\hatn) + i X_{(l2)}^*(\hatn)].
\label{partialtwo}
\end{equation}
Upon summing and differencing and carrying out the integration
over $\phi$, we obtain
\begin{equation}
     C_l^{\rm G} = -\pi N_l^2 \int_0^\pi
     d\theta\sin\theta [C^Q(\theta) G^+_{(l2)}(\cos\theta) +
     C^U(\theta) G^-_{(l2)}(\cos\theta)],
\label{Gtransform}
\end{equation}
\begin{equation}
     C_l^{\rm C} = -\pi N_l^2 \int_0^\pi
     d\theta\sin\theta [C^U(\theta) G^+_{(l2)}(\cos\theta) +
     C^Q(\theta) G^-_{(l2)}(\cos\theta)]
\label{Ctransform}
\end{equation}
which are the desired relations giving the polarization
multipole moments in terms of the polarization autocorrelation
functions.
Given some measured correlation functions, evaluation of
Eq.~(\ref{Ctransform}) for any $l$ will probe the existence of
non-scalar modes.

\subsection{Correlation Functions in the Small-Angle Limit}

In order to make contact with previous work and to present
estimates useful for measurements on a small patch of the sky,
we now derive the small-angle limit of the above expressions which give 
correlation functions in terms of multipole moments (and {\it
vice versa}).  The correlation functions given in previous work
were of Stokes parameters measured in a
fixed coordinate basis, whereas
ours are of Stokes parameters measured with respect to the great
arc connecting the two points being correlated.  However, our
results can be compared with previous results by taking $\phi=0$
in their expressions.   Although the expressions for correlation
functions in the small-angle limit given in
Refs.~\cite{annals,flatsky} are quite complicated, when
the small-angle limit is taken consistently in all steps, the
expressions simplify greatly, as emphasized in
Ref.~\cite{seljak96} (and resemble correlation functions for
ellipticities of galaxies due to weak lensing from large-scale
inhomogeneities \cite{albert,kaiser}).

Once again we begin with the temperature moments. A useful
asymptotic relation is
\begin{equation}
     P_l(\cos\theta) \sim J_0(s), \qquad 
     s\equiv(2l+1)\sin(\theta/2) \rightarrow 0,
\label{Plapprox}
\end{equation}
where $J_m(s)$ is the Bessel function of order $m$. 
Substituting into Eq. (\ref{TTresult}),
approximating the sum by an integral and taking the
limit $l\gg1$ gives
\begin{equation}
     C^{\rm T}(\theta) \simeq {1\over 2\pi} \int^\infty\, l\, dl\,
     J_0(l\theta) \, C_l^{\rm T}
\label{TTsmall}
\end{equation}
for $\theta\ll1$.  For the temperature-polarization
cross-correlation function, we note that
\begin{equation}
     P_l^2(\cos\theta) \sim 4 l^4\, J_2(s), \qquad 
     s\rightarrow0,
\label{Pltwoapprox}
\end{equation}
which gives
\begin{equation}
     C^{\rm TQ}(\theta) \simeq {2^{3/2} \over \pi} \, \int l^3
     dl\, C_l^{\rm TG}\, J_2(l\theta),
\label{TQsmall}
\end{equation}
for $\theta\ll1$, from Eq. (\ref{TQresult}).

For the polarization autocorrelation functions, note that \cite{albert}
\begin{equation}
     G_{(lm)}^\pm(\cos\theta) \sim {1\over4} l^4 [J_0(s) \pm
     J_4(s)], \qquad s\rightarrow0.
\label{Gapprox}
\end{equation}
{}From Eq. (\ref{QQresult}) we obtain
\begin{equation}
     C^Q(\theta) \simeq - { 1 \over 2\pi} \int
     l\, dl\, [(C_l^{\rm G}+C_l^{\rm C})J_0(l\theta) + 
               (C_l^{\rm G}-C_l^{\rm C})J_4(l\theta)],
\label{QQsmall}
\end{equation}
and from Eq. (\ref{UUresult}) we obtain
\begin{equation}
     C^U(\theta) \simeq - {1 \over 2\pi} \int
     l\, dl\, [(C_l^{\rm G}+C_l^{\rm C})J_0(l\theta) - 
               (C_l^{\rm G}-C_l^{\rm C})J_4(l\theta)],
\label{UUsmall}
\end{equation}
for $\theta \ll 1$.

Eqs. (\ref{TTsmall}), (\ref{TQsmall}), (\ref{QQsmall}), and
(\ref{UUsmall}) agree with the forms in Eq.~(19) in
Ref.~\cite{seljak96} for $\phi=0$ and
$C_l^{\rm C}=0$.  (Also recall that $\VEV{Q_r
U_r}=\VEV{TU_r}=0$.)  If $C_l^{\rm C}=0$, then $\VEV{Q_r Q_r +
U_r U_r}$ (which depends on an integral over $J_0(l\theta)$) and
$\VEV{Q_r Q_r - U_r U_r}$ (which depends on an integral over
$J_4(l\theta)$) depend on the same set of moments $C_l^{\rm G}$
and are therefore {\it not} independent.  However, if $C_l^{\rm
C} \neq 0$, then these correlation functions will depend on two
independent sets of moments.  

We can also derive expressions for the multipole moments for
$l\gg1$ in terms of a correlation function measured at small
angular separations.  For example, using
Eq. (\ref{Plapprox}) to approximate Eq. (\ref{TfromTT}) for
$l\gg1$ gives
\begin{equation}
     C_l^{\rm T} \simeq 2 \pi \int J_0(l\theta) C^{\rm T}(\theta)
     \theta d\theta
\label{TfromTTsmall}
\end{equation}
and using Eq. (\ref{Pltwoapprox}) to approximate Eq. (\ref{TGfromTQ})
gives
\begin{equation}
     C_l^{\rm TG} \simeq 4\sqrt{2} \pi l^2 \int
     J_2(l\theta) C^{\rm TQ}(\theta) \theta d\theta.
\label{TGfromTQsmall}
\end{equation}
Using Eq. (\ref{Gapprox}), we can approximate
Eq. (\ref{Gtransform}) by
\begin{equation}
     C_l^{\rm G} \simeq -{\pi\over 2}
     \int \left( C^Q(\theta)[J_0(l\theta) +
     J_4(l\theta)] + C^U(\theta)[J_0(l\theta) - J_4(l\theta)]
     \right)\theta d\theta ,
\label{Gtransformsmall}
\end{equation}
and Eq. (\ref{Ctransform}) by
\begin{equation}
     C_l^{\rm C} \simeq -{\pi\over 2}
     \int \left( C^U(\theta)[J_0(l\theta) +
     J_4(l\theta)] + C^Q(\theta)[J_0(l\theta) - J_4(l\theta)]\right)
     \theta d\theta.
\label{Ctransformsmall}
\end{equation}
If any nonzero $C_l^{\rm C}$ is found in this way
with correlation functions measured on a small patch of the sky,
it is an indication of vector or tensor modes.

\section{Summary and Discussion}

This paper provides a detailed and complete formalism for
characterizing polarization fluctuations in a full-sky
map of the cosmic microwave background. We give explicit
forms for tensor spherical harmonics, in which the polarization
can be expanded in direct analogy to the expansion of
temperature perturbations in the usual spherical harmonics.
The tensor harmonics are numerically just as easy to evaluate
as spherical harmonics, so polarization map simulation and analysis
will be no more cumbersome than in the temperature case.

The most important physics results presented here are that
of the six sets of multipole coefficients describing the correlations
in a temperature/polarization map, two must be zero if the Universe
is parity-invariant, and a third vanishes for scalar metric
perturbations. The moments $C_l^{\rm C}$, which are non-zero
only for vector and tensor metric perturbations, are in principle
an unambiguous probe of primordial gravity waves and vorticity
\cite{letter,sz96b}. A cosmological contribution to the moments
$C_l^{\rm TC}$ or $C_l^{\rm CG}$ would demonstrate a remarkable
handedness to the primordial perturbation spectrum. A much more
likely and practical use of these moments is to monitor foreground
microwave emission. We also note that CMB polarization may give
useful information on primordial magnetic fields
\cite{kl96,bfield} and galaxy cluster magnetic fields
\cite{inprepkl}.  Measurement of polarization in the
Sunyaev-Zeldovich effect can be used to measure cluster
transverse velocities \cite{polarsz} and/or the CMB quadrupole
moment incident on the cluster \cite{kamloeb}.

Most current microwave background codes calculate the Legendre
coefficients of the radiation brightness in Fourier space \cite{codes}. 
We have derived exact expressions for all of the multipole moments
in terms of these brightness coefficients. For tensor metric
perturbations, the expressions are particularly simple and
trivial to implement numerically. The result for scalar perturbations
is somewhat more complex, involving an infinite sum over the
brightness moments. However, the contribution of the
sum to the final expression for the multipole moments is only
significant for the lowest moments, so the overall cost of the
computation should only increase slightly. The formulas for
the multipole moments derived in this paper should allow for
relatively simple conversion of existing CMB codes.

Of course, a cosmological signal will have to be distinguished
{}from foreground contamination.  
Synchrotron emission from our galaxy is
highly polarized \cite{synchrotron}, and extragalactic radio sources
may also contribute significantly \cite{radiosources}.
The amplitude of these foreground polarization
sources is unknown at the present time.  Since both
likely foregrounds have a spectral dependence substantially
different from the blackbody CMB spectrum, the usual techniques for
subtracting foregrounds from temperature maps should also work
for polarization \cite{foregrounds}.  

Of course, simply attaining the necessary sensitivity to make
any polarization detection will be a great experimental
accomplishment. The MAP satellite, currently being constructed,
will have the sensitivity to make a statistical detection of
polarization. The COBRAS/SAMBA satellite, now in the planning
stage, should be capable of seeing polarization
on a pixel-by-pixel basis if it is configured to measure
polarization. At this time, it is undecided whether
COBRAS/SAMBA, which uses incoherent bolometer detectors
in its most sensitive frequency channels, 
will sacrifice some temperature sensitivity to
make polarized measurements. But optimistically, within a decade
we may have in hand detailed temperature/polarization maps of
the cosmic microwave background. How much cosmological information
can be extracted from such maps is currently under 
study \cite{paramnext}. The
formalism presented in this paper provides a basis for addressing such
questions.

\acknowledgments

We would like to thank Jerry Jungman, Lloyd Knox, and Uros
Seljak for helpful conversations, and
David Spergel and Gary Hinshaw for stimulating questions.
This work was supported by D.O.E. contract DEFG02-92-ER
40699, NASA NAG5-3091, and the Alfred P. Sloan Foundation at
Columbia, NASA AST94-19400 at FNAL, and the Harvard Society of
Fellows.  M.K. acknowledges the hospitality of the
NASA/Fermilab Astrophysics Center and the CERN Theory Group.

\appendix
\section{Differential Geometry on the Sphere}

This appendix collects results from differential geometry, with particular
application to the manifold $S^2$ (the 2-sphere), which are needed in
definitions of and calculations with the tensor spherical harmonics on the
celestial sphere.  We use the notation $f_{,a}\equiv{\partial f/\partial
x^a}$ to indicate a regular partial derivative and 
$f_{:a}\equiv\nabla_a f$ for a covariant
derivative. We use the colon, ``:'', rather than the more
traditional semi-colon, ``;'', to distinguish derivatives on $S^2$ from
4-dimensional derivatives in general relativity. All of
our tensors are defined with respect to a coordinate basis. Note
the the metric tensor $g_{ab}$
commutes with covariant differentiation: $g_{ab:c}=0$.
The determinant of the metric is denoted by $g\equiv ||g_{ab}||$.
Covariant derivatives of scalars, vectors, and (rank-2) tensors are
\begin{equation}
S_{:a}=S_{,a} \qquad
V^a{}_{:b}=V^a{}_{,b}+V^c\Gamma^a_{bc} \qquad
T^{ab}{}_{:c}=T^{ab}{}_{,c}+T^{db}\Gamma^a_{cd}+T^{ad}\Gamma^b_{cd}
\end{equation}
where the $\Gamma^a_{bc}$ are Christoffel symbols defined by
\begin{equation}
\Gamma^a_{bc}={1\over2}\,g^{ad}(g_{db,c}+g_{dc,b}-g_{bc,d}) .
\end{equation}
We have used the formula for the 4th derivatives of a scalar function
\begin{equation}
S^{:ab}{}_{ab}=\nabla^2\nabla^2S+R^{db}S_{:db}+{1\over2}\,R^{:d}S_{:d} \qquad
\nabla^2S\equiv S^{:a}{}_a \qquad R_{ab}\equiv R^c{}_{acb} \qquad 
R\equiv R^a{}_a
\end{equation}
where $R_{abcd}$, $R_{ab}$, and $R$ are the Riemann tensor, the Ricci tensor,
and Ricci scalar, respectively. We have also used the formula of integration by
parts
\begin{equation}
\oint d^2\hatn \sqrt{g} X^{ab} Y_{:ab} 
= -\oint d^2\hatn \sqrt{g} X^{ab}{}_{:a} Y_{:b}
= \oint d^2\hatn \sqrt{g} X^{ab}{}_{:ab} Y,
\label{byparts}
\end{equation}
where $\oint$ indicates integration over a closed manifold with no boundary,
and $\hatn$ is shorthand for $dx^1 dx^2$.

	On a two dimensional manifold the Levi-Civita symbol is a traceless
anti-symmetric rank-2 tensor given by
\begin{equation}
\epsilon_{ab}=\sqrt{g}\,\left(\matrix{ 0 & 1 \cr
                                      -1 & 0 }\right)
\end{equation}
which has the following simple properties:
\begin{equation}
\epsilon_{ca}\epsilon^c{}_b=g_{ab}=-\epsilon_{ac}\epsilon^c{}_b, \qquad\qquad
\epsilon_{ab}\epsilon_{cd}  =g_{ac}g_{bd}-g_{ad} g_{bc},         \qquad\qquad
\epsilon_{ab:c}=0 .
\label{epsproperties}
\end{equation}
A rank-2 tensor is trace-free (or traceless) iff $g^{ab}T_{ab}=0$, and a rank-2
tensor is symmetric iff $T_{ab}=T_{ba}$.  In two dimensions the latter
requirement is equivalent to $\epsilon^{ab}T_{ab}=0$.  Linear polarization is
described by a symmetric trace-free rank-2 tensor, or STF for short.

	The geometry of a two dimensional manifold is particularly simple
because it is determined solely by the Ricci scalar $R$.  Some handy
identities are
\begin{equation}
              R_{abcd}={1\over2}\,R\,\epsilon_{ab}\epsilon_{cd} \qquad
                R_{ab}={1\over2}\,R\,g_{ab}                    \qquad
 \epsilon^{ab}R_{abcd}=           R\,\epsilon_{cd}             \qquad
 \epsilon^{ac}R_{abcd}={1\over2}\,R\,\epsilon_{bd} .
\end{equation}
Another useful identity is
\begin{equation}
M^{ab}N^{cd}\epsilon_{ac}\epsilon_{bd}=-M^{ab}N_{ab} \qquad {\rm if} \qquad
g^{ab}M_{ab}=g^{ab}N_{ab}=\epsilon^{ab}M_{ab}=\epsilon^{ab}N_{ab}=0,
\label{ShotNoiseIdentity}
\end{equation}
i.e. for STF tensors only.

	In this paper we are only interested in the geometry of the unit
sphere, $S^2$.  Its geometry is nearly as simple as can be, since $R=2$.  We
exclusively use the explicit coordinate system defined by
spherical polar coordinates,
$(\theta,\phi)$, where $\theta$ is the polar angle from a
particular point on the sphere, and $\phi$ labels the angle on circles which
are centered on this same point.  In these coordinates the metric is
\begin{equation}
g_{ab}=\left(\matrix{1 & 0\cr
                     0 & \sin^2\theta\cr}\right),
\qquad\qquad g\equiv\Vert g_{ab}\Vert=\sin^2\theta.
\label{metric}
\end{equation}
while the antisymmetric tensor is
\begin{equation}
\epsilon_{ab} = \left(\matrix{0 & \sin\theta \cr
                              -\sin\theta & 0\cr}\right),
\qquad\qquad
\epsilon^a{}_b = \left(\matrix{0 & \sin\theta \cr
                               -\csc\theta & 0 \cr}\right).
\label{epstensor}
\end{equation}
{}From this metric the Christoffel symbols follow as
\begin{equation}
\Gamma^\theta_{\phi\phi}= -\sin\theta\cos\theta \qquad
\Gamma^\phi_{\theta\phi}=\Gamma^{\phi}_{\phi\theta}=\cot\theta \qquad 
  \Gamma^\theta_{\theta\theta} 
= \Gamma^\theta_{\theta\phi  }
= \Gamma^\theta_{\phi  \theta}
= \Gamma^\phi_{  \theta\theta}
= \Gamma^\phi_{  \phi  \phi  }=0.
\label{christoffel}
\end{equation}
The explicit components of the 2nd derivatives of a scalar function are
\begin{eqnarray}
Y_{:ab} &=& Y_{,ab} - \Gamma^c_{ab} Y_{,c}\,;\nonumber\\
Y_{:\theta\theta} &=& Y_{,\theta\theta}\,,\nonumber\\
Y_{:\theta\phi} &=& Y_{,\theta\phi}
    -\cot\theta\, Y_{,\phi}\,,\nonumber\\
Y_{:\phi\phi} &=& Y_{,\phi\phi}
    +\sin\theta\cos\theta\, Y_{,\theta},
\label{covderivs}
\end{eqnarray}
while explicit expressions for the some second 
derivatives of a symmetric rank-2 tensor are
\begin{eqnarray}
     M^{ab}{}_{:ab} = M^{\theta\theta}{}_{,\theta\theta} + 
     2 M^{\theta\phi}{}_{,\theta\phi} + 
     M^{\phi\phi}{}_{,\phi\phi}
     & - &\sin\theta\cos\theta M^{\phi\phi}{}_{,\theta}
     \nonumber\\ 
     & + &2 \cot\theta  M^{\theta\theta}{}_{,\theta}
     + 4 \cot\theta   M^{\theta\phi}{}_{,\phi}
     + (1-3\cos^2\theta) M^{\phi\phi} - M^{\theta\theta}
\label{MderivsG}
\end{eqnarray}
and
\begin{eqnarray}
     M^{ab}{}_{:ac}\epsilon^c{}_b  = 
     \sin\theta\left(M^{\theta\phi}{}_{,\theta\theta}
                     + M^{\phi\phi}{}_{,\phi\theta}\right)
     & - &\csc\theta \left(M^{\theta\theta}{}_{,\theta\phi}
                     + M^{\phi\theta}{}_{,\phi\phi}\right)
       -\cot\theta\csc\theta M^{\theta\theta}{}_{,\phi}\nonumber\\
     &&\qquad\qquad + 5\cos\theta M^{\theta\phi}{}_{,\theta}
                    + 3\cos\theta M^{\phi\phi}{}_{,\phi}
                    + 3\left(\cos\theta\cot\theta -
		    \sin\theta\right) M^{\theta\phi},
\label{MderivsC}
\end{eqnarray}
which again only apply if $\epsilon^{ab}M_{ab}=0$.

\section{Spherical Harmonic and Legendre Function Identities}

In this Appendix, we list some identities involving spherical
harmonics and Legendre functions which we have used in our
calculations.

The associated Legendre functions are defined by
\begin{equation}
     P^m_l(x) = (-1)^m(1-x^2)^{m/2} {d^m\over dx^m}P_l(x),
      \qquad\qquad m\geq 0,
\label{pmldef}
\end{equation}
\begin{equation}
P^{-m}_l(x) = (-1)^m {(l-m)!\over(l+m)!}P_l^m.
\label{pmlnegdef}
\end{equation}
Normalization:
\begin{equation}
\int_{-1}^1 dx P^m_l(x) P^m_{l'}(x) = \delta_{ll'}{2\over 2l+1}
{(l+m)!\over(l-m)!}
\label{plmnorm}
\end{equation}
Legendre Polynomials:
\begin{equation}
P_l(x)\equiv P_l^0(x).
\end{equation}

The associated Legendre functions satisfy the following
recursion relations (see Eqs. 8.733 in
Ref.~\cite{grad}):
\begin{equation}
  P^{m+2}_l(x) +2(m+1){x\over\sqrt{1-x^2}}P^{m+1}_l(x)
               + (l-m)(l+m+1)P^m_l(x) = 0,
\label{Plmrecursion}
\end{equation}
\begin{equation}
     (2l+1)\sqrt{1-x^2}P^{m-1}_l(x) = P^m_{l-1}(x) - P^m_{l+1}(x),
\label{plmidentity2}
\end{equation}
\begin{equation}
(2l+1)xP_l^m = (l-m+1)P_{l+1}^m + (l+m)P_{l-1}^m,
\label{plmidentity1}
\end{equation}
\begin{equation}
(1-x^2){dP_l^m\over dx} = (m+1)xP_l^m - (l-m+1)P^m_{l+1},
\label{dplmdx1}
\end{equation}
\begin{equation}
(1-x^2){dP_l^m\over dx} = -lxP_l^m + (l+m)P^m_{l-1}.
\label{dplmdx2}
\end{equation}

Using orthonormality and completeness of Legendre polynomials, an
associated Legendre function of $m=2$ can be written in terms of
Legendre polynomials using the integral \cite{coulsonthesis}
\begin{equation}
\int_{-1}^1 dx P_l(x) P_j^2(x) = \cases{0,&$l>j$ or $j+l$ odd;\cr
        -2j(j-1)/(2j+1),& $j=l$;\cr
        4,& $l<j$ and $j+l$ even\cr}
\label{plpl2int}
\end{equation}

With the above conventions for Legendre functions, the
spherical harmonics are given by
\begin{equation}
Y_{(lm)}(\theta,\phi) = \sqrt{{2l+1\over 4\pi}{(l-m)!\over(l+m)!}}
    P_l^m(\cos\theta) e^{im\phi}.
\label{ylmdef}
\end{equation}
The convolution of a Legendre polynomial with a
spherical harmonic is
\begin{equation}
     \int d\hatn\, P_j(\hatk \cdot \hatn) \, Y_{(lm)}(\hatn) =
     {4\pi \over 2l+1} Y_{(lm)}(\hatk) \, \delta_{lj}.
\label{PYconvolution}
\end{equation}
This can be obtained by expressing the Legendre polynomial in
terms of spherical harmonics with the spherical harmonic
addition theorem and then using the orthonormality of spherical
harmonics. 

Angular-momentum lowering and raising operators can be used to
derive the following recursion relations for spherical harmonics:
\begin{equation}
     e^{i\phi}\sin\theta Y_{(lm)} = - \left[ {(l+m+1)(l+m+2)
     \over (2l+1)(2l+3)}\right]^{1/2} Y_{(l+1,m+1)} 
     + \left[ {(l-m)(l-m-1)
     \over (2l-1)(2l+1)}\right]^{1/2} Y_{(l-1,m+1)},
\label{Ylmrecurrenceone}
\end{equation}
\begin{equation}
     e^{-i\phi}\sin\theta Y_{(lm)} = \left[ {(l-m+1)(l-m+2)
     \over (2l+1)(2l+3)}\right]^{1/2} Y_{(l+1,m-1)} 
     - \left[ {(l+m)(l+m-1)
     \over (2l-1)(2l+1)}\right]^{1/2} Y_{(l-1,m-1)},
\label{Ylmrecurrencetwo}
\end{equation}
which can be iterated and evaluated for $m=0$ to obtain
\begin{eqnarray}
     \cos2\phi\sin^2\theta Y_{(l0)} = {1\over2}&&\Biggl\{ \left[
     {(l+1)(l+2)(l+3)(l+4) \over (2l+1)(2l+3)^2 (2l+5) }
     \right]^{1/2} (Y_{(l+2,2)}+Y_{(l+2,-2)}) \nonumber \\
     & - & 2 {\sqrt{ l (l-1)(l+1)(l+2)} \over (2l-1)(2l+3)}
     (Y_{(l,2)}+Y_{(l,-2)}) \nonumber \\
     & + & \left[ {l(l-1)(l-2)(l-3) \over (2l-3)(2l-1)^2(2l+1)}
     \right]^{1/2} (Y_{(l-2,2)}+Y_{(l-2,-2)}) \Biggr\}.
\label{iteratedone}
\end{eqnarray}
For the replacement $\cos2\phi \rightarrow \sin2\phi$
make the replacements $(1/2) \rightarrow (1/2i)$ in the
prefactor and $(Y_{(x,2)}+Y_{(x,-2)}) \rightarrow
(Y_{(x,2)}-Y_{(x,-2)})$.

Looking at the $\theta$ dependence of Eq. (\ref{iteratedone}),
we get
\begin{equation}
     (1-x^2) P_j(x) = {P_{j+2}^2 \over (2j+1)(2j+3)} - {2 P_j^2 \over
     (2j-1)(2j+3) } + {P_{j-2}^2\over (2j+1)(2j-1)}.
\label{parttwo}
\end{equation}

\end{document}